\documentclass[namedreferences]{solarphysics}
\usepackage[optionalrh]{spr-sola-addons} 
\usepackage{graphicx}        
\usepackage{color}           
\usepackage{url}             

\begin{document}

\begin{article}

\begin{opening}

\title{Simulation of $f$-Mode Propagation Through a Cluster of Small Identical Magnetic Flux Tubes\\ {\it Solar Physics}}

\author{K.~\surname{Daiffallah}$^{1}$
       }
\runningauthor{K.Daiffallah}
\runningtitle{$f$-Mode Interaction with a Cluster of Magnetic Flux Tubes }

   \institute{$^{1}$ Observatory of Algiers, CRAAG, Algiers, Algeria, 
                     email: \url{k.daiffallah@craag.dz} \\ 
                          }

\begin{abstract}
Motivated by the question of how to distinguish seismically between 
monolithic and cluster models of sunspots, we have simulated the
propagation of an $f$-mode wave packet through two identical small
magnetic flux tubes ($R=200$ km), embedded in a stratified 
atmosphere. We want to study the effect of separation $d$ and incidence angle
$\chi$ on the scattered wave. We have demonstrated that the horizontal
compact pair of 
tubes ($d/R=2$, $\chi=0$) oscillate as a single tube when the incident wave is propagating, which gives a scattered wave amplitude of about twice that from a single tube. The scattered amplitude decreases
with increasing $d$ when $d$ is about $\lambda/2\pi$ where $\lambda$ is the wavelength of the incident wave packet. In
this case the individual tubes start to oscillate separately in the manner of  
near-field scattering. When $d$ is about twice of  $\lambda/2\pi$, scattering from individual tubes reaches the far-field regime, giving rise to coherent scattering  with an amplitude similar to the case of the compact pair of tubes. For perpendicular incidence
  ($\chi=\pi/2$), the tubes oscillate simultaneously with the incident wave packet. Moreover, simulations show that a compact cluster oscillates almost as a
  single individual small tube and acts more like a scattering object, while a loose cluster shows multiple-scattering in
  the near-field and the absorption is largest  when $d$ within the cluster is about $\lambda/2\pi$.  
This is the first step to understand the seismic response of a bundle of
magnetic flux tubes in the context of sunspot and plage helioseismology.

\end{abstract}


\end{opening}

\section{Introduction}
     \label{S-Introduction} 

One of the unsolved problems in solar physics concerns the magnetic structure of sunspots. Direct observations were unable to resolve this structure because of lack of sufficient spatial resolution, but also
because of the hidden structure of the magnetic field below the visible solar surface. 
The simplest model is a monolithic magnetic flux tube which remains more or less homogeneous with increasing depth \cite{cowling53}. 

The alternative model was proposed by \inlinecite{parker79} where the magnetic field of sunspot spreads into
 several discrete magnetic flux tubes below the visible surface of the Sun (cluster model). The debate between these two models is still relevant. However, \inlinecite{thomas82} have suggested that indirect observations
can answer the question about the structure of sunspot magnetic field. They argued that solar acoustic waves
interact with sunspots, so that it would be possible to probe the structure beneath the surface by studying the observed wave field. The observations of  Braun, Duvall, and LaBonte (\citeyear{braun87}, \citeyear{braun88})
indicate that there is a significant absorption of $f$- and $p$-modes by sunspots. From this result, it is recognized as important to demonstrate that helioseismic waves interact differently with two distinct configurations of sunspot magnetic field.  
It was established that resonant absorption by
fibril models of a sunspot can significantly increase the total acoustic energy
absorption. Furthermore, fibril models produce a significant
absorption across a wide range of plausible parameter values that can be
adjusted unlike the monolithic model \cite{rosenthal90}.
The first investigations about the interaction of acoustic waves with a
bundle of magnetic flux tubes was limited to a statistical approach where a
number of tubes are distributed uniformly or randomly in an infinite
homogeneous atmosphere \cite{ryutov76,zweibel89}.
These methods average the governing equations, which is less difficult
mathematically, but relevant physical information will be lost.

Sunspots can
also scatter waves. However, the theory of multiple scattering by magnetic flux
tubes has been ignored in the past years due to the complexity of the
problem. The multiple scattering was treated by using the formalism developed
by \inlinecite{bogdan91}. They studied the scattering of acoustic waves by a
pair of uniform magnetic flux tubes for a series of separations in an
unstratified atmosphere. Using the same formalism, \inlinecite{keppens94}
studied wave interaction with a bundles of magnetic flux tubes in different
geometrical configurations. This method finds the solution for each magnetic flux tube of the 
cluster considering both the incident acoustic wave and 
the scattered wave from all the other tubes. 
An important implication of these studies is that the cluster of tubes is more
effective in absorption of acoustic waves than individual isolated tubes. For a
single tube, the excitation is set by the incoming wave and it depends on the
ratio of the tube-radius to the wavelength of the incoming wave ($R/\lambda$). However, in the flux
tube bundle, the excitation must take into account multiple scattering of waves.

If the separation between the tubes within the cluster is not larger than the wavelength of the incoming wave, such that the tubes are in each others' near-field zones, the dominant excitation at each tube is provided by nearby
neighbors  rather than the contribution from the incident plane wave, leading to
greatly enhanced scattered wave fields (multiple scattering).
 
When the tubes are far from the others ($d \gg$ $\lambda$), the total scattering cross section is not different from the sum of individual
flux-tube scattering cross section. Generaly, the scattered fields interfere
destructively in the far field (incoherent scattering).

For an intermediate separation, the scattering cross section of the bundle
starts varying with the separation $d$, adding a coherent component to the
cross section of the bundle (coherent scattering).

An important result is that the scattering behavior of a monolithic tube cannot
be distinguished from that of a closely-packed fibril sunspot of the same
magnetic flux.

The efforts mentioned above have been limited to the unstratified atmosphere, mainly
because of the mathematical complexity that is introduced by the stratification. The
analytical treatment is even more complicated for the scattering and
absorption of waves by a cluster of magnetic flux tubes, while it is already
quite complicated for the case of a single tube.

By using the same theoretical approach of \inlinecite{hanasoge08}
to describe the scattering matrix of a single thin-flux-tube, \inlinecite{hanasoge09} studied oscillation modes and
scattering of a pair of flux tubes embedded in a gravitationally stratified
atmosphere. They found that the strongest coupling is between the $f$-mode and
the flux tubes where the dominant interaction distance is about half the
horizontal surface wavelength of the incident waves ($\approx \pi/k$). They noticed
also that the scattering coeifficients attain large values at small flux tube
separations. These results are similar to those of \inlinecite{bogdan91} and
\inlinecite{keppens94} who suggested that a pair or a bundle of magnetic flux
tubes can absorb waves quite effectively compared to a single monolithic tube. 

We simulate in this study the propagation of a linear $f$-mode wave packet through a bundle of magnetic flux
tubes in a three-dimensional polytropic stratified atmosphere. The aim is to
distinguish between the monolithic and the cluster models of sunspots by
studying the scattered waves and understand the interaction between waves and plage. The paper is organized as follows. In
Section 2,  we briefly describe the numerical code that we used and the set up
of the simulations. In Section 3,  we present the results of simulations of
wave propagation through a pair of magnetic flux tubes which are arranged in parallel or perpendicular to the incident wave packet. In Sections 4 and 5, we compare the
scattering of a monolithic magnetic flux tube with the scattering of a compact cluster of seven tubes, and a loose cluster
of seven and nine tubes, respectively. Finally we conclude in Section 6.

\section{Simulations}
     \label{S-simulations} 

We used the {\sf{SLiM}} code \cite{cameron07} to solve the
three-dimensional linearized wave equations in Cartesian geometry defined by
the horizontal coordinates $x$ and $y$, and the
vertical coordinate $z$.
We have periodic boundary conditions on the horizontal 
side walls of the simulation box. A pseudo-spectral scheme is implemented
in the horizontal directions and a two-step
Lax-Wendroff scheme in the vertical direction to evolve the horizontal
Fourier modes.

As in \inlinecite{daiffallah11}, the horizontal domain is $x$  $\in [-20,20]$ Mm and  $y$  $\in [-10,10]$ Mm. The height
range is from 0.2 Mm to 6 Mm below the solar surface ($z$ increases with depth).
The initial condition is an $f$-mode wave packet propagating with a Gaussian envelope centered
at the angular frequency 3 mHz with standard deviation of 1.18 mHz. At $t = t_0$, the
wave packet is located at the left edge of the computational
domain $x_0 = -20$ Mm, and it propagates from left to right
in the $x$-direction. 
The background atmosphere is an enhanced polytropic atmosphere \cite{cally97}.

The initial individual magnetic flux tube is taken vertical in the $z$-direction and axisymmetiric with a radial profile given by  $ B(r) = B_{0} \exp(- r^4/ R^4) $ where $R$ 
is the tube radius, and $B_{0}$ = 4820 gauss (G). The flux tube is almost evacuated and it is superposed on the background atmosphere.
The plasma-$\beta$ value changes with depth according to the variation of the sound speed in the atmosphere. The sound speed is set to be equal to the Alfv\'en speed at a depth of 400 km.

The scattered wave field  is constructed as the difference
between the simulations with and without the flux tube.

\section{$f$-Mode Interaction with Two Identical Magnetic Flux Tubes }
     \label{pair-simulations} 
\begin{figure}  
\centering
\includegraphics[width=0.75\textwidth]{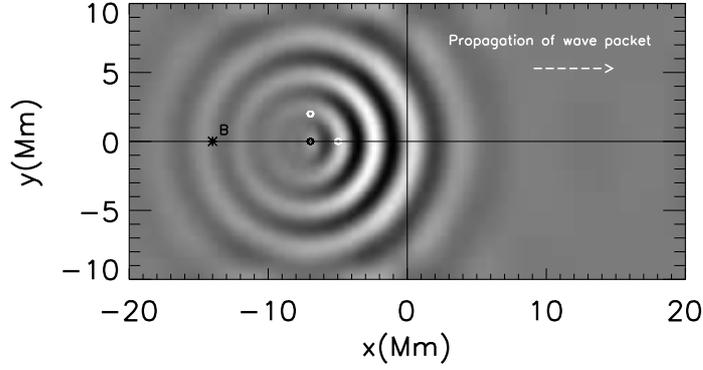} 
 \caption {Scattered wave field due to the reference tube (black contour) and the second tube (white contour). The second tube to the right
and in the $y$-direction corresponds to the angle $\chi=0$ and $\chi=\pi/2$, respectively. The separation distance $d$ between the reference tube and the
second tube changes from $d= 0.08 \lambda$ to $d = 0.4 \lambda$; $\lambda \approx $ 4.85 Mm is the wavelength of the $f$-mode. The unit tube has a radius $R=200$ km. The scattered component $V_z$  is measured at point B$(-14,0)$ for all simulations. The reference tube is situated at the position $(-7,0)$.} 
\label{pedago2}
\end{figure}

To understand the interaction of a wave with an ensemble of magnetic flux tubes, it is
useful to study the basic case where the bundle is composed of a pair
of tubes aligned  parallel or perpendicular  (in the $x-y$ plane) to the direction of propagation of the 
incident wave packet. To simplify the interpretation of results, we consider
that all individual tubes are identical with radius $R=200$ km.
Figure \ref{pedago2} shows how the pair of flux tubes are positioned in
horizontal or perpendicular configuration. The reference tube in black contour is 
situated at the position $(-7,0)$. The white contours represent successive
positions of the second tube.

We define $\chi$ as the angle between the direction of propagation of the
incident wave packet and the line connecting the reference tube to the second
tube. The second tube located on the right-hand-side of the reference tube in Figure \ref{pedago2}
corresponds to  $\chi=0$. The tube located in the
$y$-direction corresponds to  $\chi=\pi/2$. The separation between the successive 
centers of the pair of tubes varies as $d=2R, 3R, 4R, 5R$, and $10R$ which correspond to $d=0.08 \lambda, 0.12 \lambda, 0.16 \lambda, 0.2 \lambda$, and $0.4 \lambda$, respectively, where $\lambda \approx 4.85$ Mm is the wavelength of the $f$-mode ($2\pi g/\omega^2$) for the wave packet centered at 3 mHz.

We have analyzed the velocity $V_z$ at point B(-14.0) situated in the scattered wave field on the left of the reference tube.
The scattering there is more interesting to analyze because it shows oscillations of the magnetic flux tube without contribution from the
incident wave packet.

The scattering
process is predominantly restricted to $f-f$ mode, while the $f-p$ mode
conversions are very weak \cite{hanasoge08,daiffallah11}.

\subsection{A Pair of Magnetic Flux Tubes with $\chi=0$}
\begin{figure}  
\centering
\includegraphics[width=0.8\textwidth]{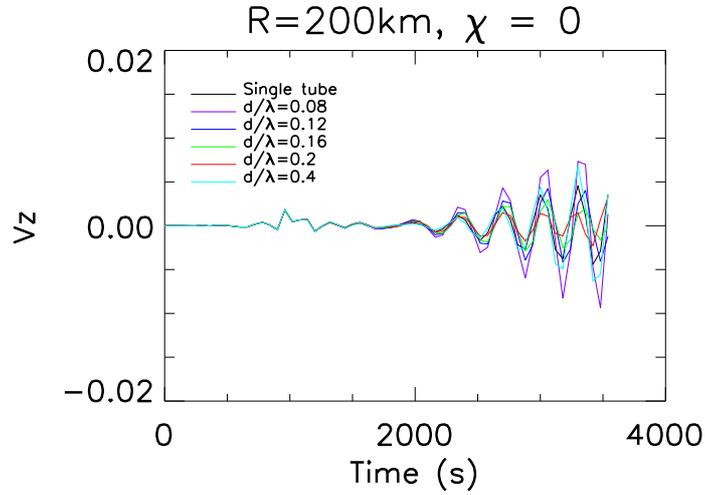} 
 \caption { Scattered vertical velocity ($V_z$) as a function of time, measured at point B
   for a pair of magnetic flux tubes ($\chi=0$). Point B is
   indicated in Figure \ref{pedago2}. The color curves are for different
   separation distances $d$ between the reference and the second tubes.} 
\label{righttime2}
\end{figure}

In this subsection, we investigate the $f$-mode interaction with a pair of magnetic flux tubes aligned in the $x$-direction ($\chi=0$). Figure \ref{righttime2} shows the time variations of 
the vertical velocity $V_z$ measured at point B as a function of the 
separation distance $d$ between the pair of flux tubes. The black
curve shows the scattering from a monolithic flux tube of 200 km radius. 

First, we note that the scattering curves of different pairs measured at point B are slightly out of phase, which means that the motion of the pair of tubes changes with the
separation $d$. The curves show that the compact pair ($d/\lambda = 0.08$) has the
largest scattering amplitude compared to the others. This amplitude is
approximatively the scattering amplitude of a single tube of 400 km radius ($2R$).

The scattering amplitude for $d/\lambda = 0.4$ is also large as the case of $d/\lambda=0.08$. 
The curves corresponding to $d/\lambda$ = 0.12, 0.16, and 0.2 are below
the curve for the compact pair $d/\lambda=0.08$.

\begin{figure} 
\centering 
\includegraphics[width=0.45\textwidth]{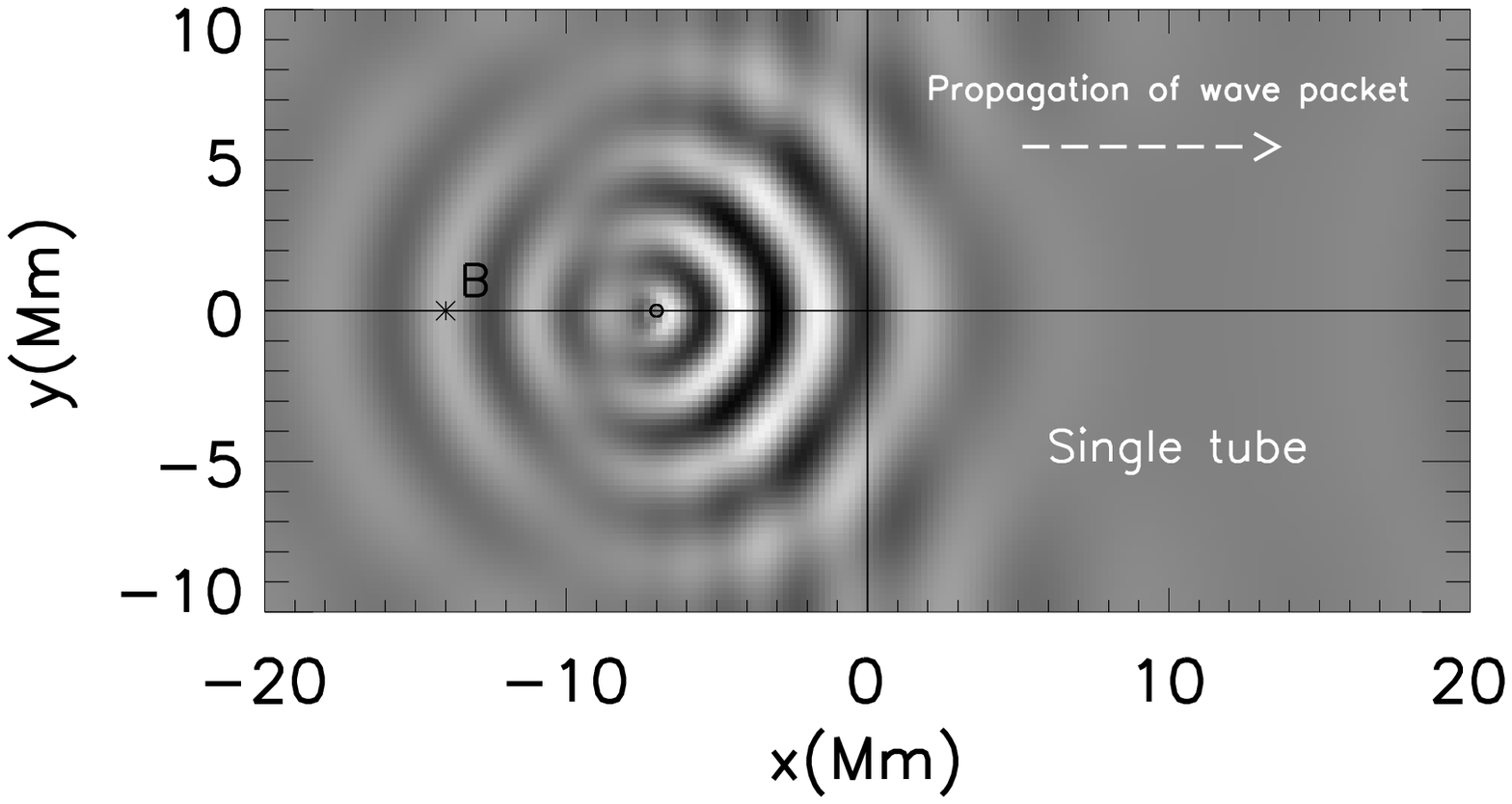} 
\includegraphics[width=0.45\textwidth]{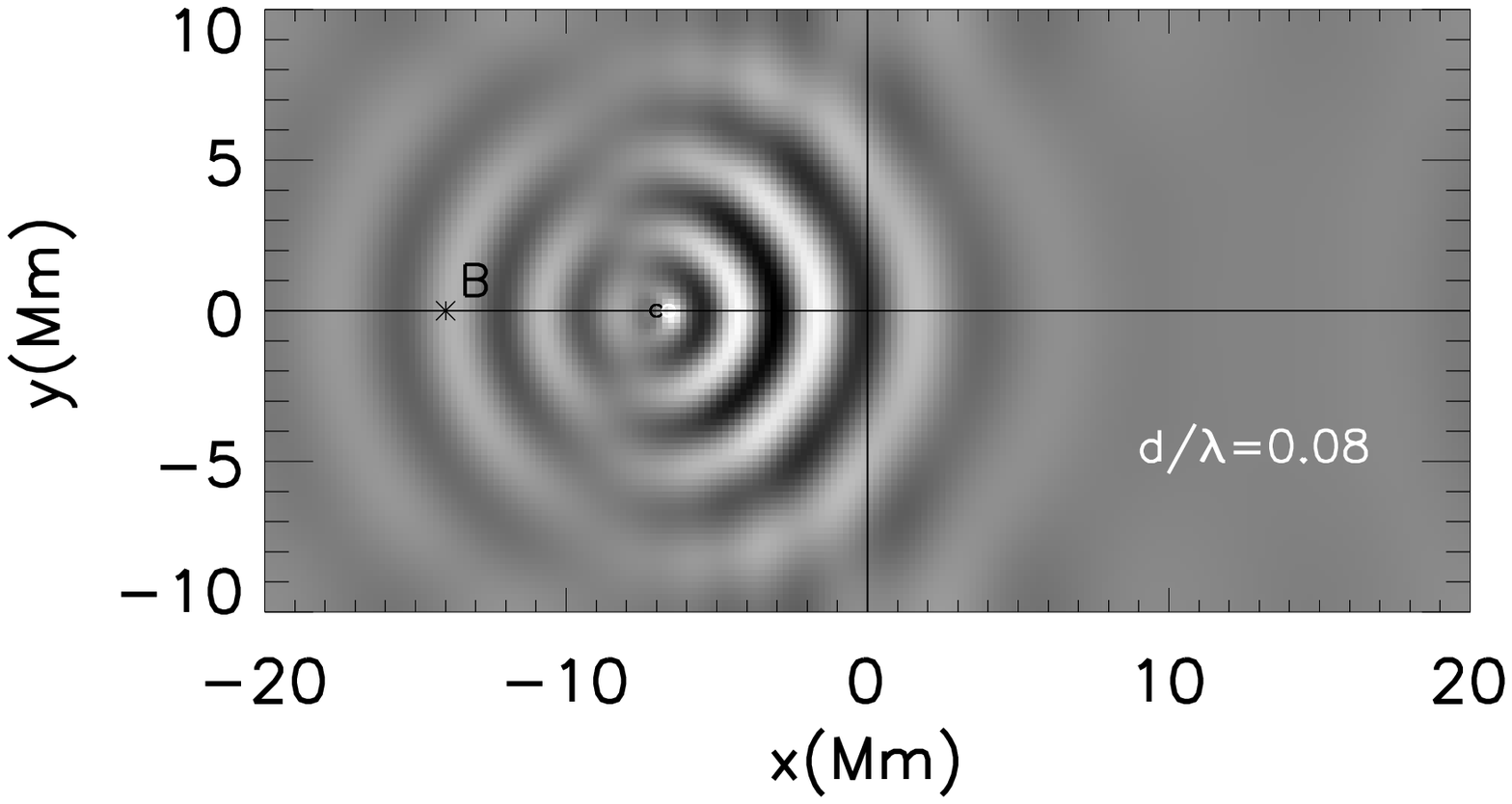} 
\includegraphics[width=0.45\textwidth]{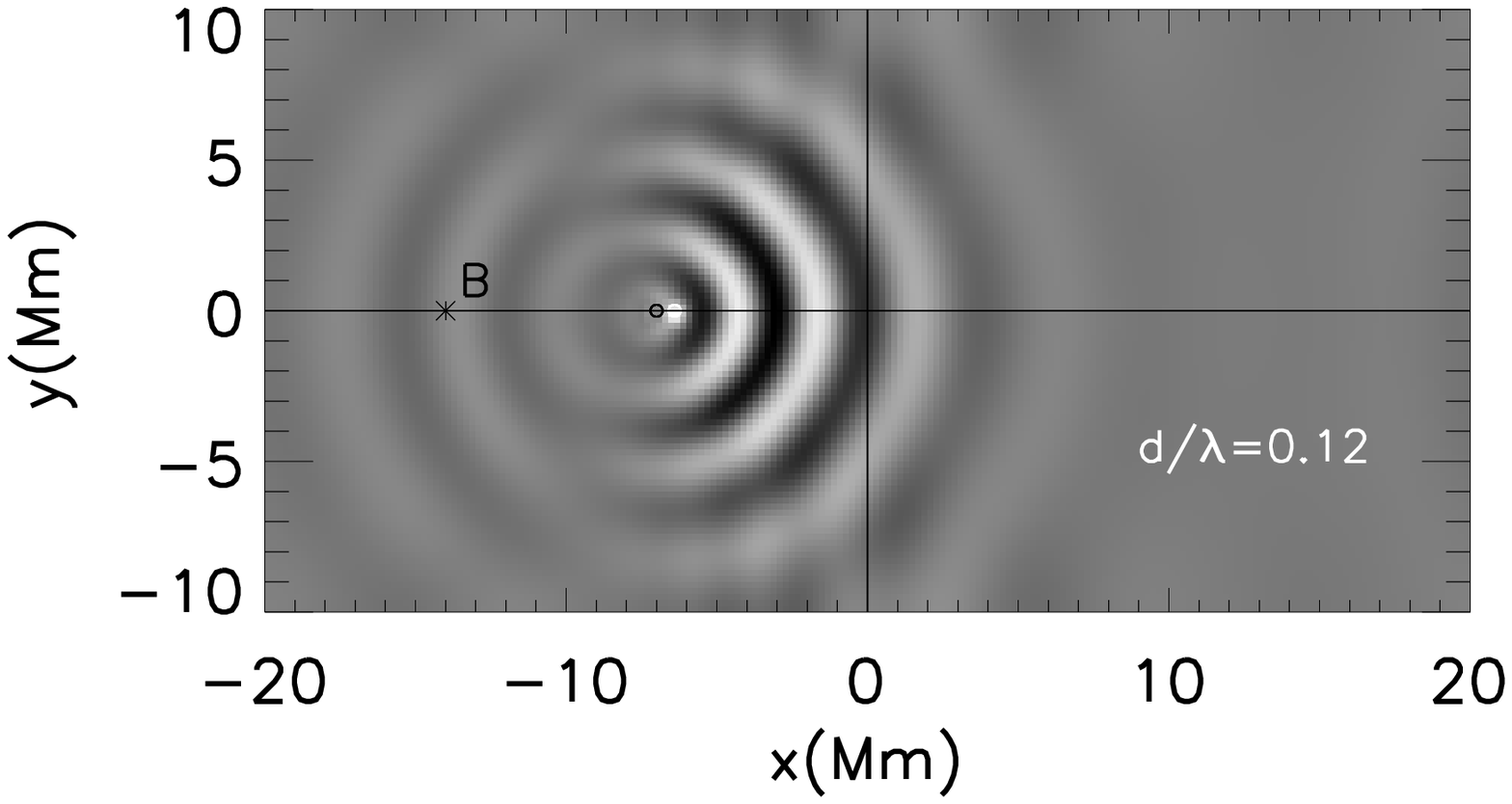} 
\includegraphics[width=0.45\textwidth]{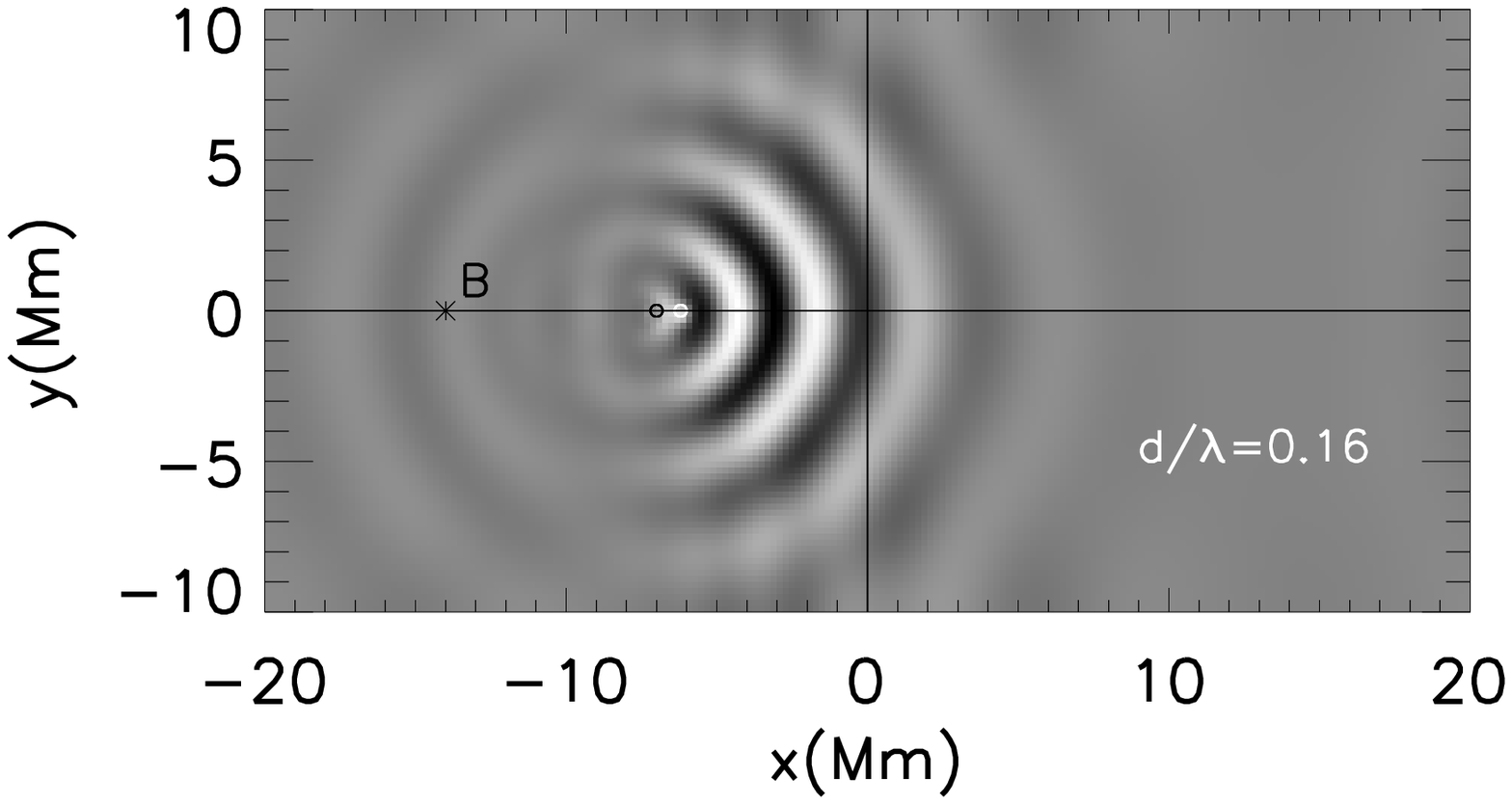} 
\includegraphics[width=0.45\textwidth]{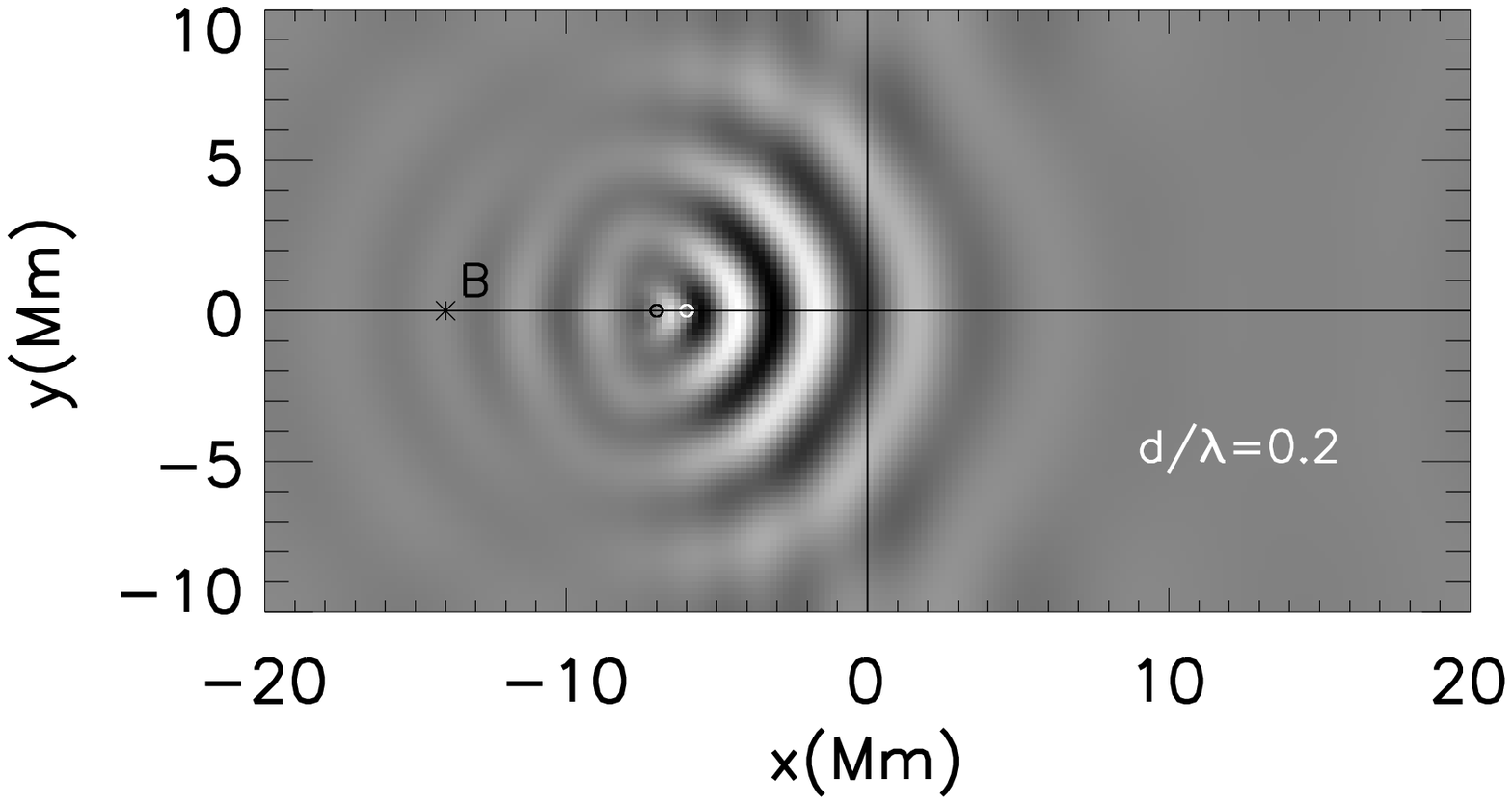} 
\includegraphics[width=0.45\textwidth]{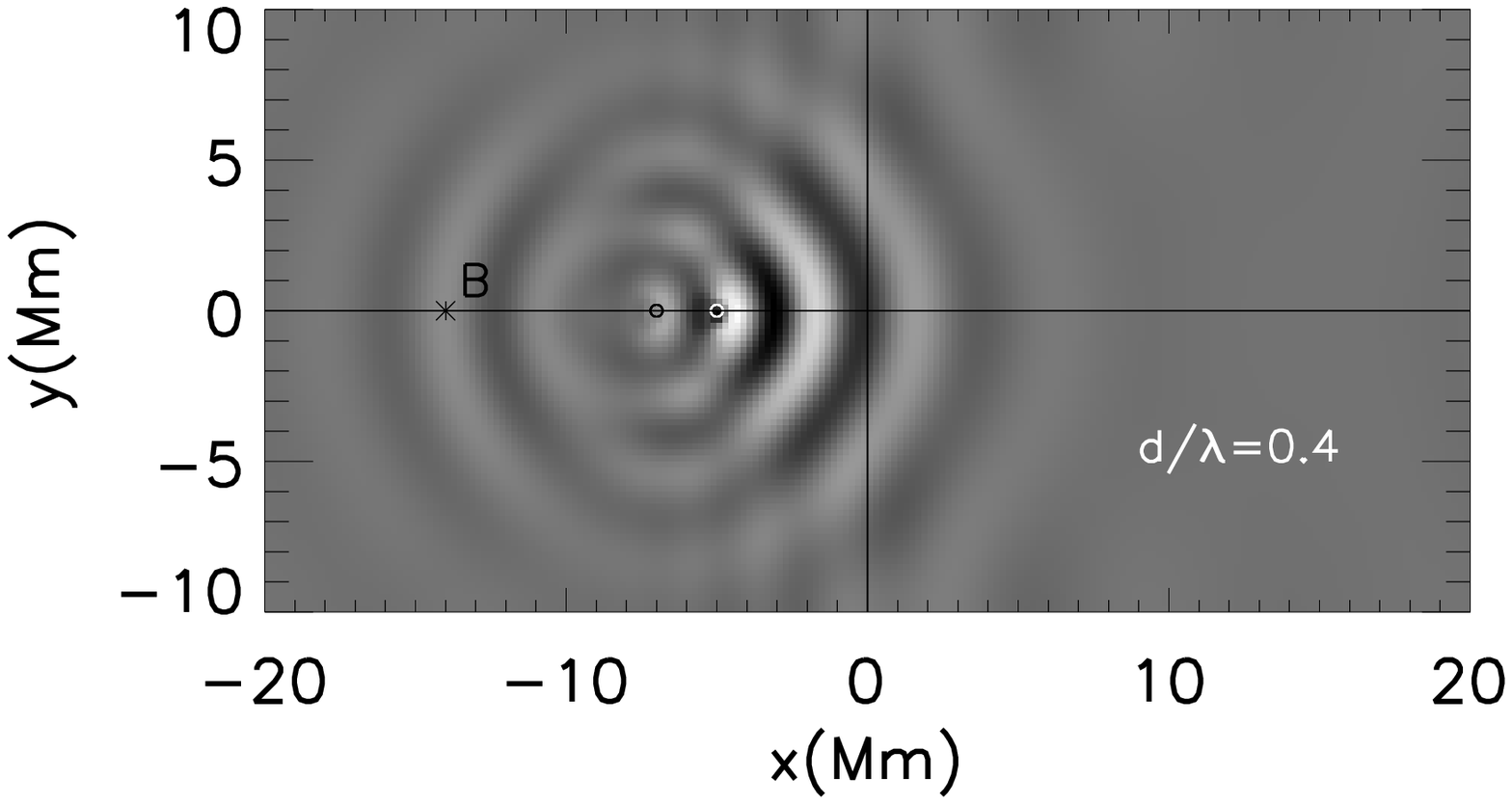} 
 \caption {Scattered wave field of $V_z$ at  $t$ = 3300 s for 
a pair of magnetic tubes with $\chi=0$. The reference and the second
tubes are represented with black and white contours, respectively. The top left figure
corresponds to the scattering of a single tube ($R=200$ km). The following figures show the
variation in scattering with the distance between the two tubes. The
separation varies as $d/\lambda = 0.08, 0.12, 0.16, 0.2$, and $0.4$, respectively.} 
\label{figrightall}
\end{figure}

Figure \ref{figrightall} shows the scattered wave field due to a pair of flux tubes with $\chi =0$ at $t=3300$ s. We can see that the wave field to the left is totally influenced by the tube oscillations. For a  separation of $d/\lambda =
0.08$, the scattered wave field to the left (dipole oscillations) is very similar
to that of a single tube, which indicates a strong coupling between the two
tubes. For the other separations, the pattern of the near-field scattering 
 changes according to the mutual interaction between the tubes. For
$d/\lambda=0.4$, we begin to clearly distinguish the waves scattered by the
tubes separately.

\begin{figure}
\centering
\vspace{0.03\textwidth}
\centerline{\bf      
     \hspace{0.3\textwidth}  \color{black}{Propagation of wave packet}
     \hfill}
\vspace{0.01\textwidth}
\centerline{\bf      
     \hspace{0.4\textwidth}  \color{black}{$------>$}
     \hfill}
\vspace{0.02\textwidth}

\includegraphics[width=0.45\textwidth]{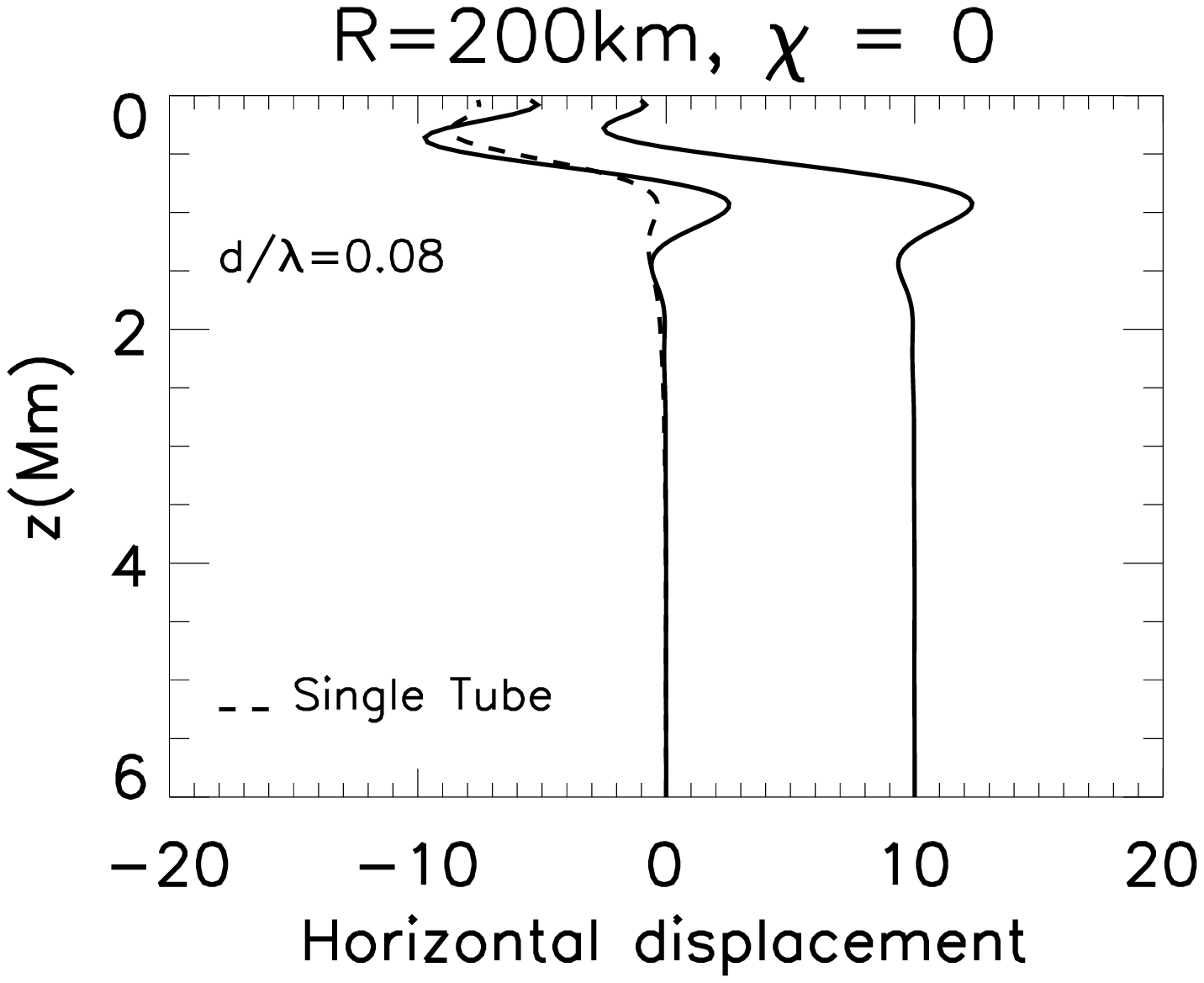} 
\includegraphics[width=0.45\textwidth]{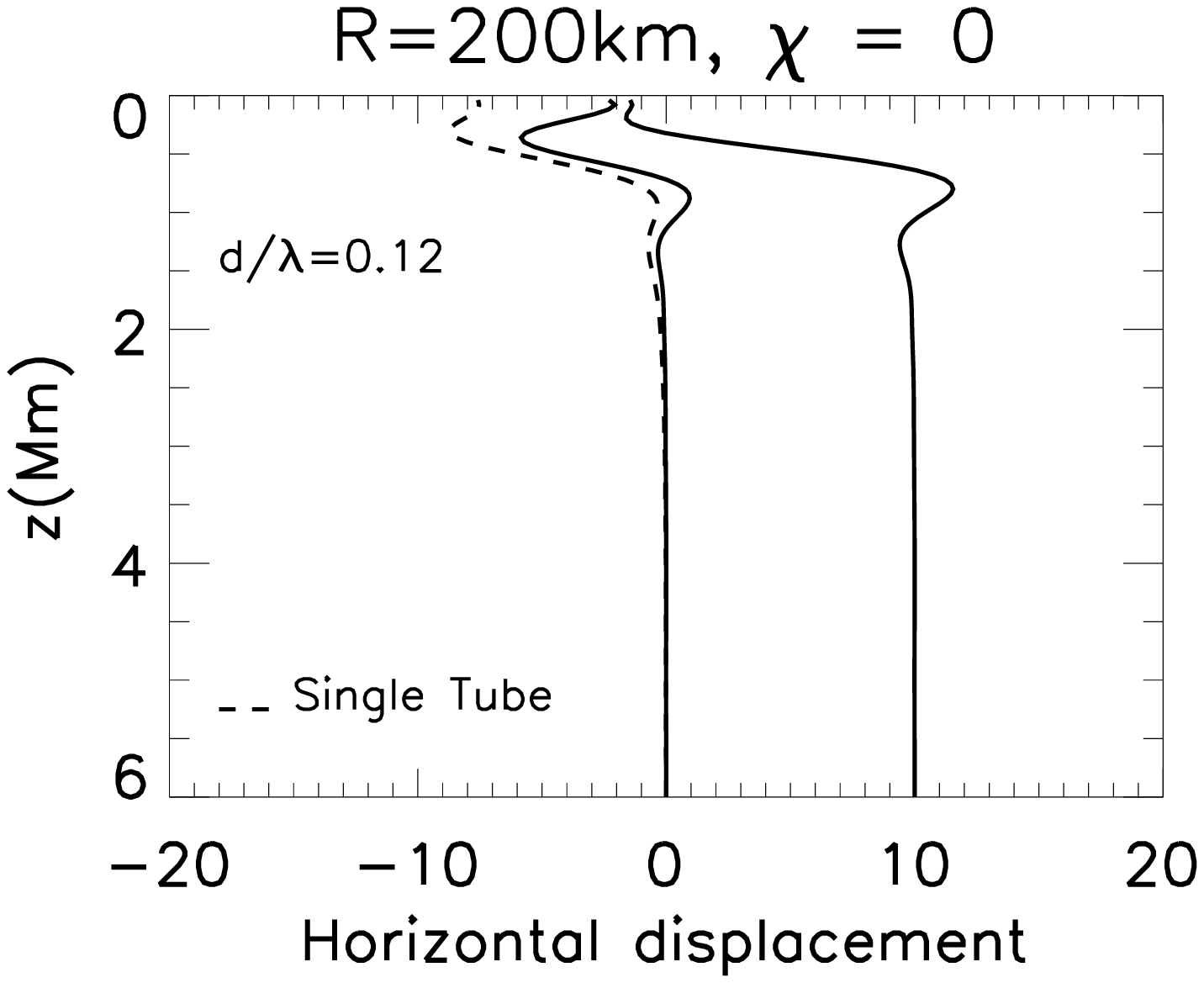} 
\includegraphics[width=0.45\textwidth]{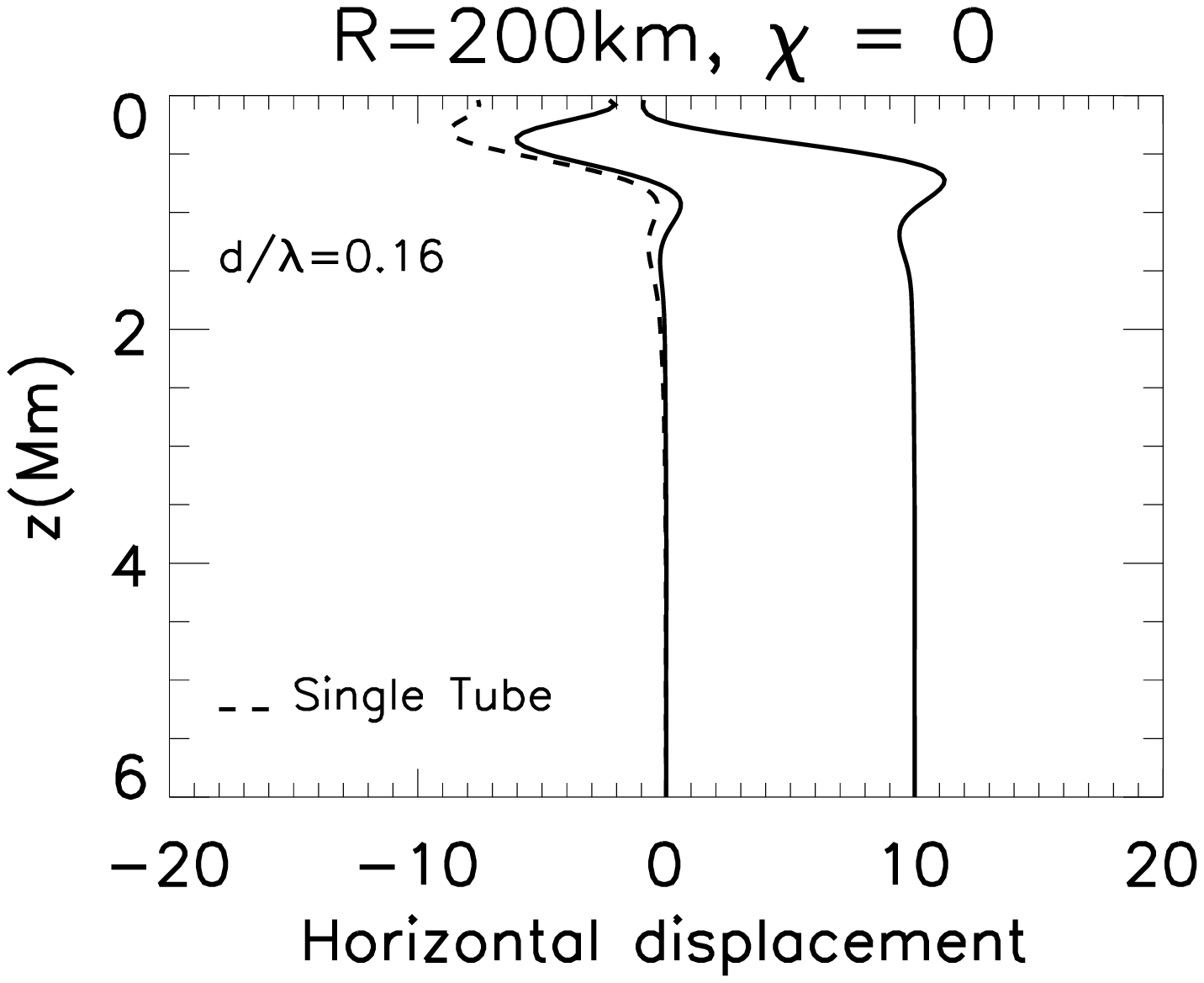} 
\includegraphics[width=0.45\textwidth]{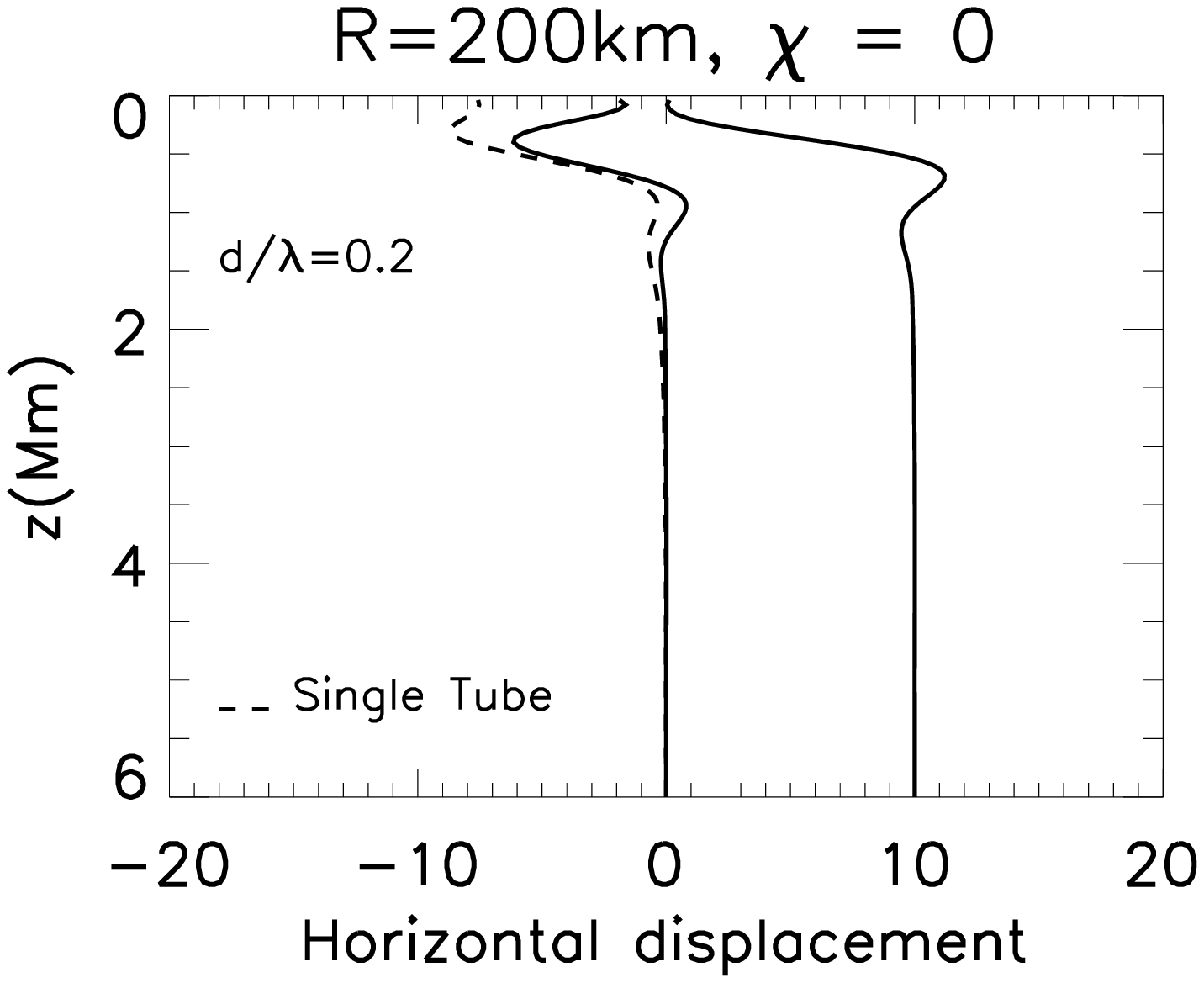} 
\includegraphics[width=0.45\textwidth]{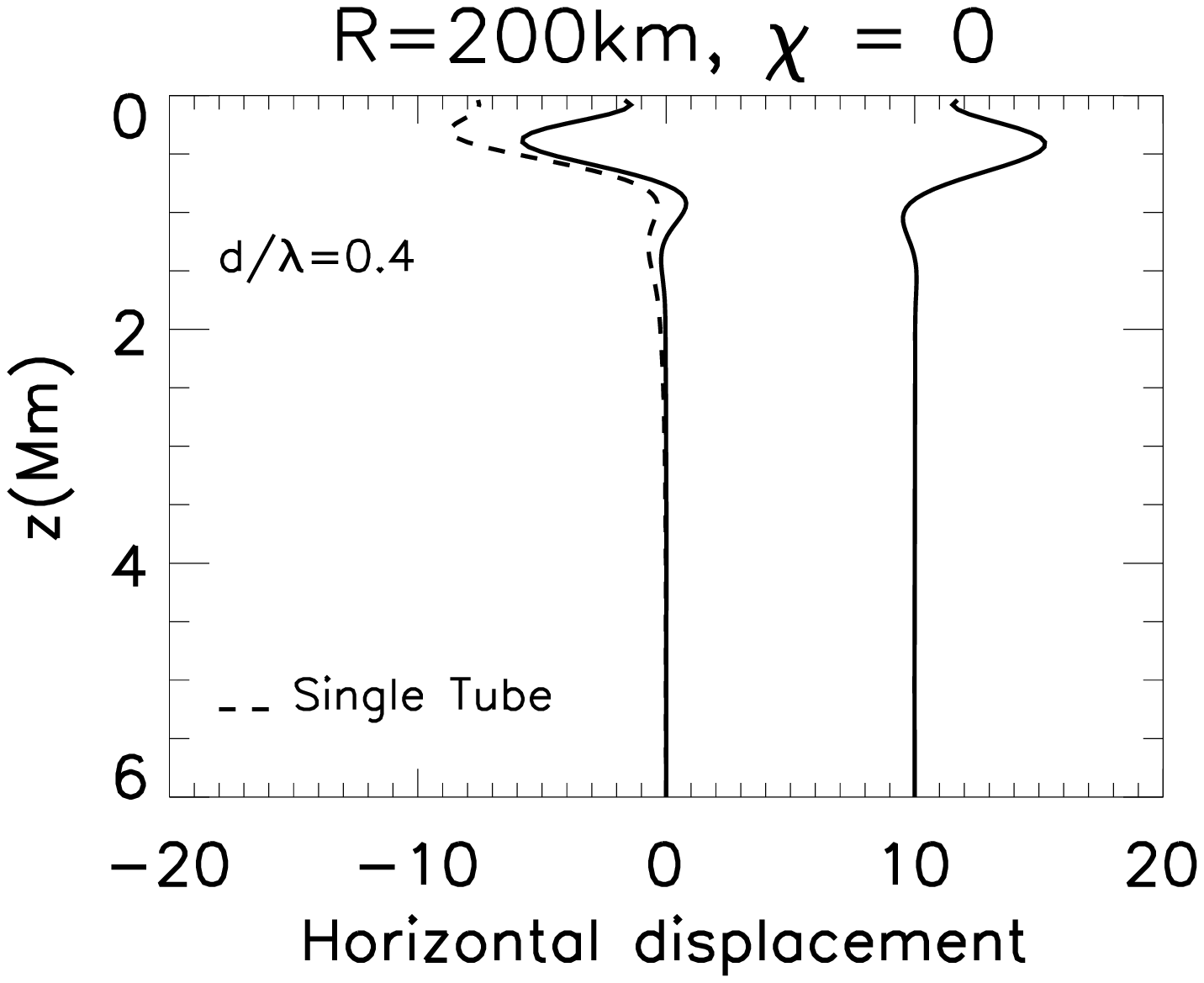} 
 \caption {Displacement of the two tube axes in arbitrary units in the $x$-direction (solide
   lines) as a function of depth $z$. The dashed curves show the $x$-displacement of
   the reference tube when it is single ($R$ = 200 km). The solid curves are snapshots at $t$ =
   2100 s after the start of the simulation and show 
   oscillations of a pair of tubes ($\chi = 0$) for various values of the
   separation  $d$.} 
\label{oscillrightallz}
\end{figure}

It is interesting to plot the $x$-displacement of the tube axis as a function of depth $z$ to see the tube oscillations. Figure \ref{oscillrightallz} shows snapshots of the $x$-displacement at $t$ = 2100 s. The solid curves are for
the pair of tubes. The dashed curve shows the oscillation of the reference
tube ($R=200$ km) when it is single. The first snapshot shows that the displacement
amplitude of the reference tube is larger than the displacement
amplitude of this
tube when it is single. This demonstrates the effect of the second tube and
consequently the increase in the amplitude when the pair is
compact. Because of the very close separation, we see that the oscillations of 
the two tubes are substantially in phase relative to the other cases with larger values of $d/\lambda$. The last snapshot for $d/\lambda= 0.4$ shows that the $x$-displacement of the tubes is completely in the opposite directions.

\subsubsection{Interference Effect between the Two Tubes with $\chi=0$}
\label{inductionsect}
\begin{figure} 
\centering 
\vspace{0.03\textwidth}
\includegraphics[width=1\textwidth]{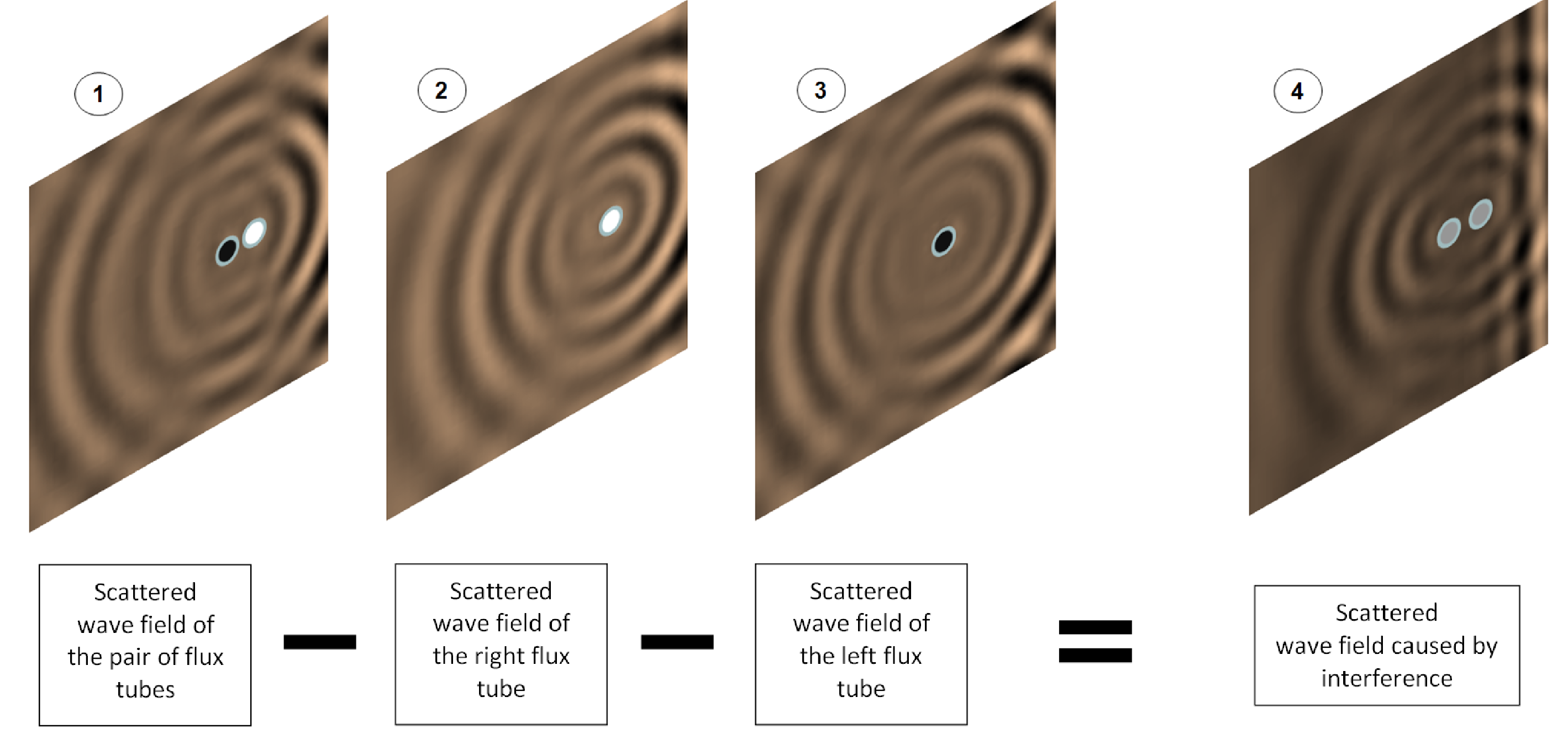} 
\vspace{0.05\textwidth}
 \caption {The interference wave field (4) is obtained by the subtraction from the image of the
scattered wave field of the two tubes (1) the image of the scattered wave field of
the reference tube (2) and the second tube (3), respectively, when they are single. The resulting
image corresponds to the effect of mutual interference of the waves from the two tubes only.}
\label{inductube}
\end{figure}

In Figure \ref{figrightall}, the scattered wave field generated by the
pair of tubes tends to obscure the mutual interference between the two tubes. This interference
is very important to understand the degree of influence of one tube to the
other depending on the separation distance $d/\lambda$. We have applied a method
to extract the total scattered field to see only the field due to interference. Figure \ref{inductube} illustrates this procedure. 

\begin{figure} 
\centering 
\vspace{0.03\textwidth}
\centerline{\bf      
     \hspace{0.3\textwidth}  \color{black}{Propagation of wave packet}
     \hfill}
\vspace{0.01\textwidth}
\centerline{\bf      
     \hspace{0.4\textwidth}  \color{black}{$------>$}
     \hfill}
\vspace{0.08\textwidth}
\includegraphics[width=0.32\textwidth]{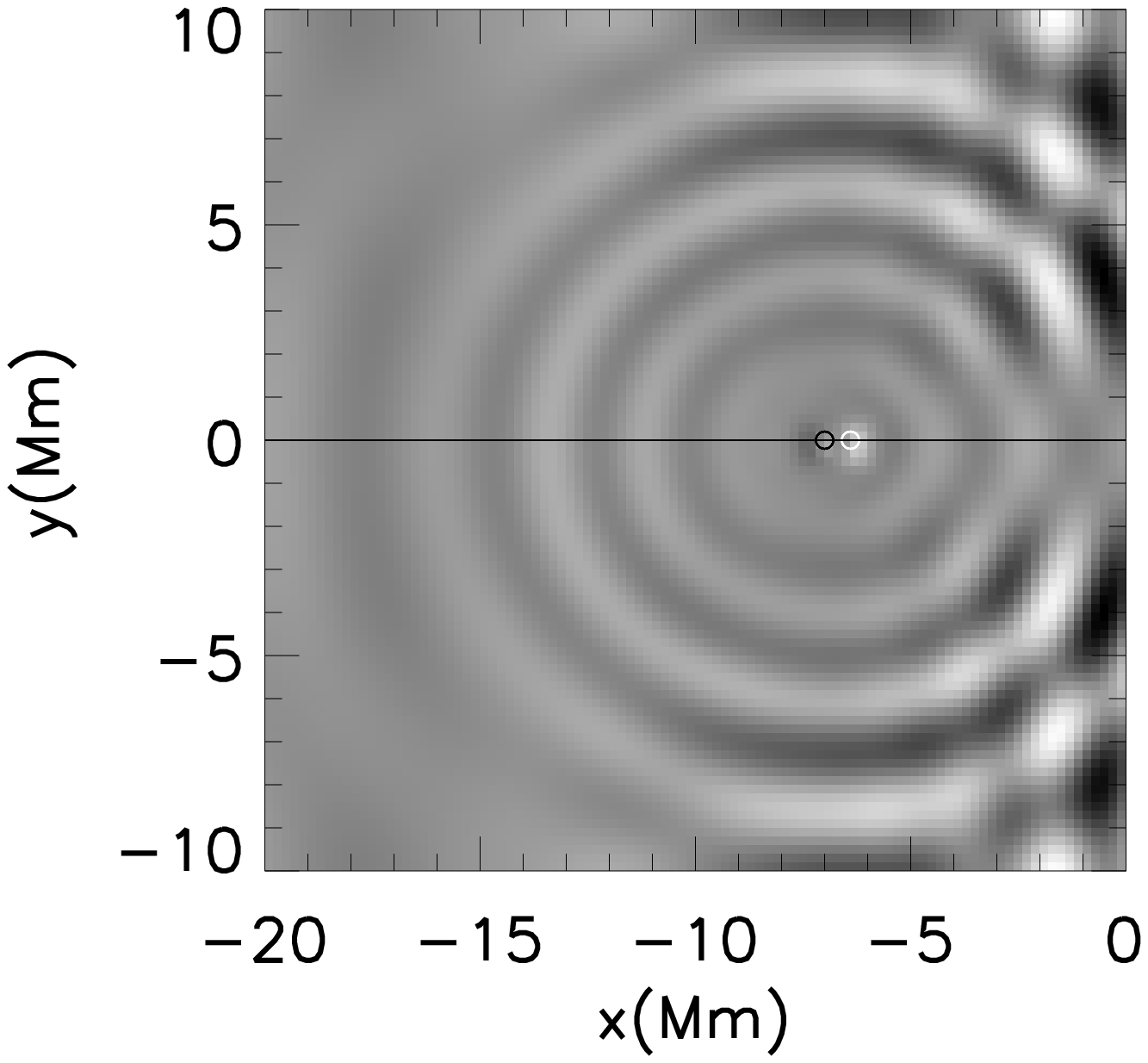} 
\includegraphics[width=0.32\textwidth]{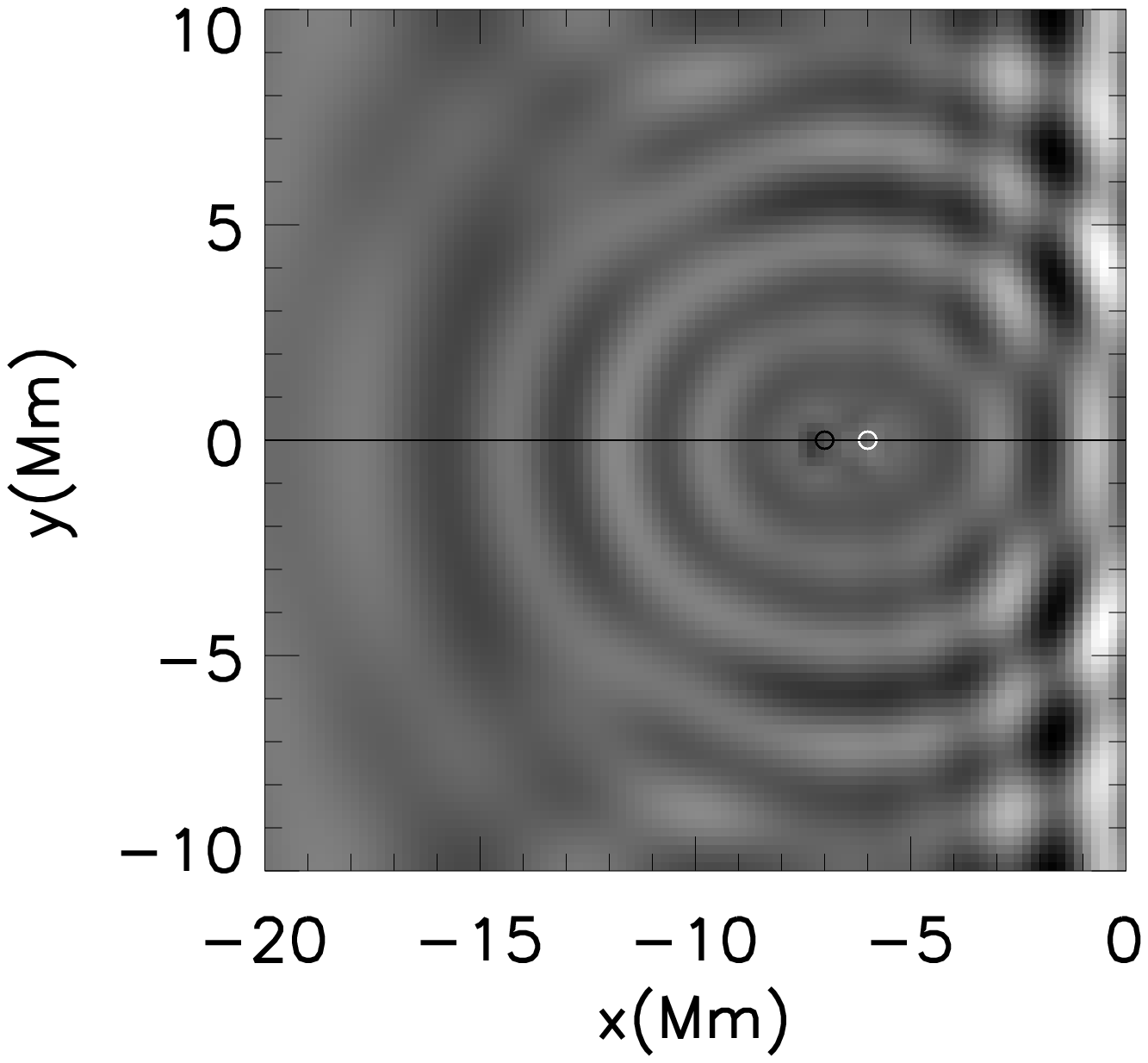} 
\includegraphics[width=0.32\textwidth]{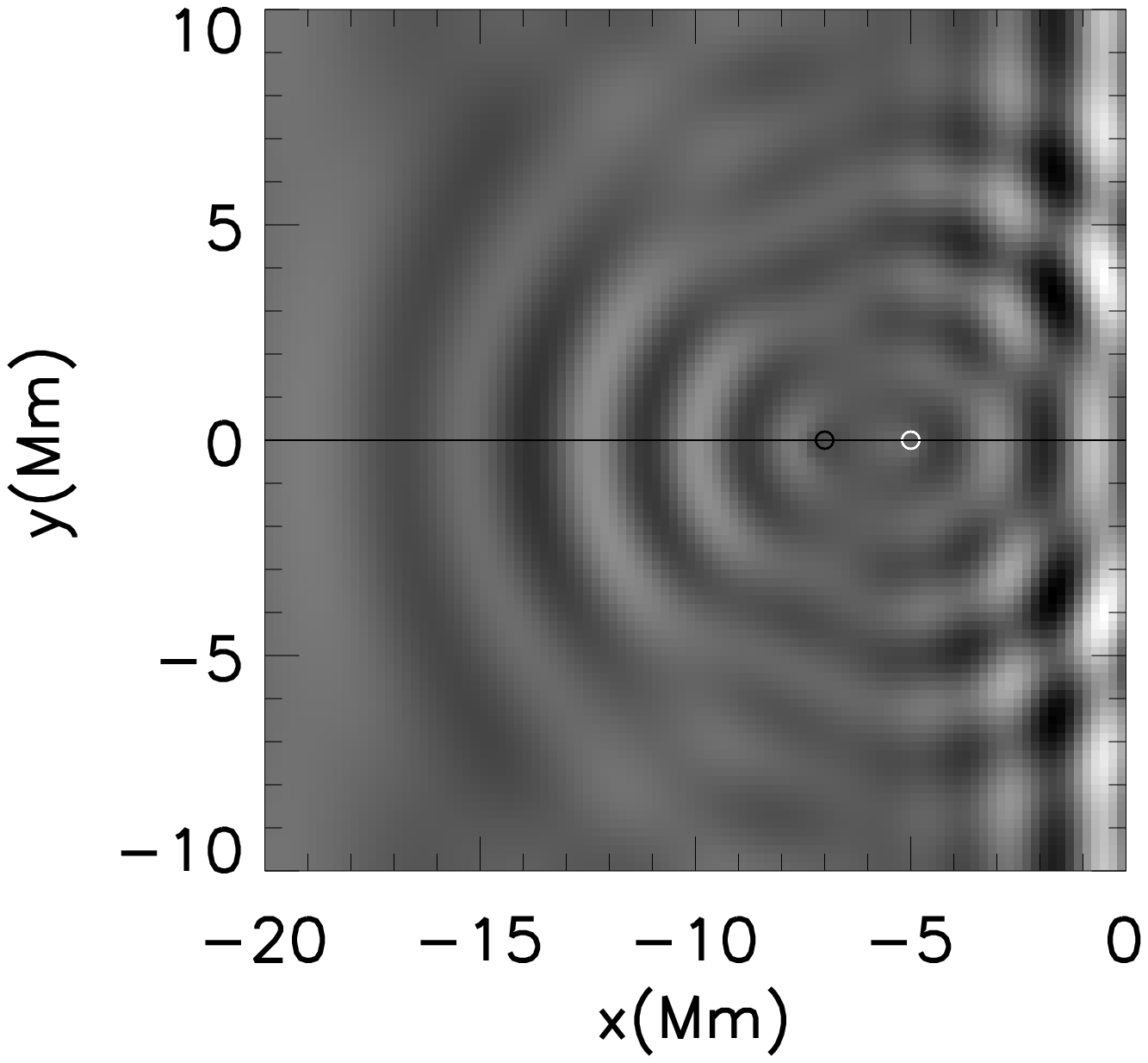} 
\vspace{0.05\textwidth}

\vspace{-0.33\textwidth}   
     \centerline{\bf      
      \hspace{0.09\textwidth}  \color{black}{$d/\lambda=0.12$}
      \hspace{0.17\textwidth}  \color{black}{$d/\lambda=0.2$}
      \hspace{0.19\textwidth}  \color{black}{$d/\lambda=0.4$}
         \hfill}
     \vspace{0.3\textwidth}

 \caption {Snapshots at $t=4200$ s of the vertical component $V_z$ showing the interference field of the two tubes
   ($\chi$ = 0) for various values of the separation $d$. From left to right $d/\lambda = 0.12, 0.2$,  and 0.4, respectively.}
\label{coupeinduitxyhoriz}
\end{figure}

This method gives rise to some noisy behavior of the wave field to the right of the tubes, but this will not disturb our analysis since we are interested in the wave field to the left of the tubes.
Figure \ref{coupeinduitxyhoriz} shows the interference field of velocity $V_z$ for pairs with separation  $d =
0.12 \lambda, 0.2 \lambda$, and $0.4 \lambda$, respectively. 
In these snapshots, waves scattered by the reference tube to the left are the result of waves
scattered by the second tube and vice versa. In the case of $d/\lambda = 0.12$, the tubes are
so close that their oscillations are almost in phase. As a result, the interference effect
produced by the tubes remains the same during the propagation of the incident
wave, which explains the symmetrical pattern of the waves scattered by the
tubes on both sides (left and right wave fields). 
The mutual interference between the tubes starts to be different as
the separation $d$ increases. In the case of $d/\lambda = 0.4$, the interference field is
no more symmetrical since the tubes oscillate differently.

We have found that the amplitude of mutual excitation of tubes in the $x$-direction 
is minimum for small separation $d$ and increases with $d$. This is
in good agreement with the previous results where the amplitude of the waves
scattered by the pair of tubes decreases with increasing $d$ when $d \le
0.2\lambda$. In this case, the pair of tubes
oscillate almost as a single tube leaving only a small energy to the oscillations of
the individual tubes, which explains maximum scattering and low
interference for the case of $d=0.08 \lambda$. Thus, the change in the scattered near-field to the left of the tubes in Figure \ref{figrightall} for $ 0.12 \lambda \le d \le 0.2 \lambda$ is caused by the small oscillations of the tubes. This corresponds to the multiple scattering regime.

\begin{figure}  
\centering
\includegraphics[width=0.8\textwidth]{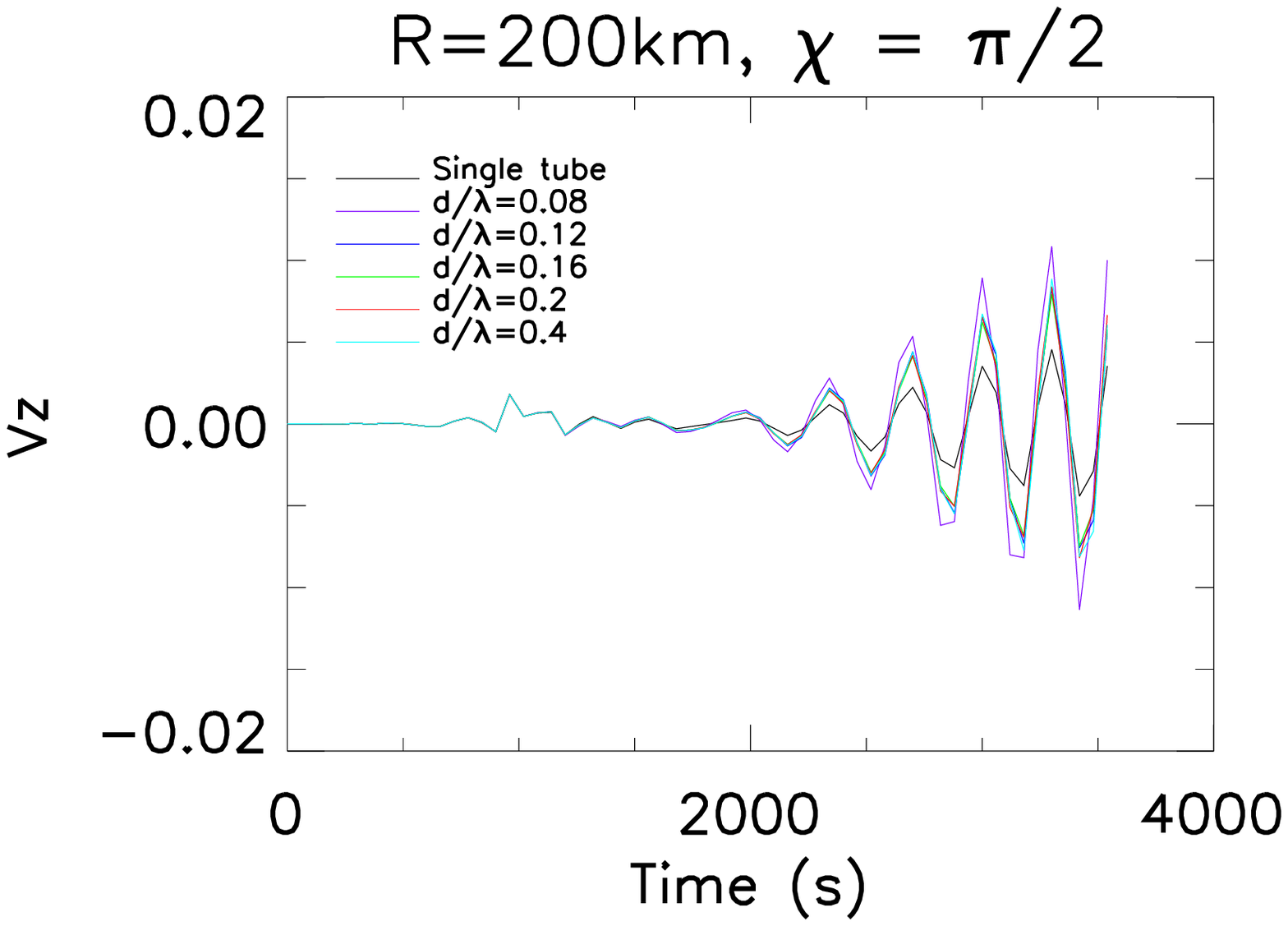} 
 \caption { Scattered vertical velocity as a function of time measured at point B
   for a pair of magnetic flux tubes ($\chi=\pi/2$). Point B is
   indicated in Figure \ref{pedago2}. The color curves are for different
   separation distances $d$ between the reference and the second tubes.} 
\label{verticaltime2}
\end{figure}

For larger separation $d > 0.2 \lambda$, the individual tubes
start to oscillate more effectively with respect to the collective oscillation of
the pair of tubes. Therefore, a part of the power of the incident wave will
supply these oscillations. Consequently, the mutual interaction
between the tubes increases. However, the scattering measured at point B for $d=0.4 \lambda$ (Figure \ref{righttime2})
increases again compared to the case of $d=0.2 \lambda$. In fact, the different phases of
 scattered waves by the tubes interact and interfere constructively in the far-field giving rise to a coherent
 scattering regime.

\subsection{A Pair of Magnetic Flux Tubes with $\chi=\pi/2$}
We simulate in this subsection the interaction of an $f$-mode wave packet with a pair
of magnetic flux tubes aligned in the $y$-direction ($\chi=\pi/2$) (Figure \ref{pedago2}). Figure
\ref{verticaltime2} shows the time variations of the scattered vertical
velocity $V_z$ measured at point B for different separation distances $d$ .

The incident wave arrives and excites simultaneously the pair of the tubes. The
consequence is that the different curves are perfectly in phase. Figure
\ref{verticaltime2} shows
that the scattered $V_z$ is maximum for compact tubes ($d/\lambda = 0.08$) as in the case
of $\chi=0$.
We can approximate this configuration of two compact
tubes aligned in the $y$-direction to a single tube of 400 km radius ($2R$), which explains
the increase in the scattered wave amplitude. However, we note that
this amplitude exceeds that of the case $\chi=0$, which implies the
contribution of both tubes in the scattering measured at point B unlike the
case of $\chi=0$.

The scattered wave field for $ d/\lambda = 0.08, 0.12, 0.16$, and 
0.2  shows similarity with the scattering from a single tube. In the case of $d/\lambda = 0.4$, the individual contribution of each tube starts to appear in the near-wave field,  but the tubes stay synchronized according to Figure \ref{verticaltime2}.

\begin{figure}  
\centering
\includegraphics[width=0.8\textwidth]{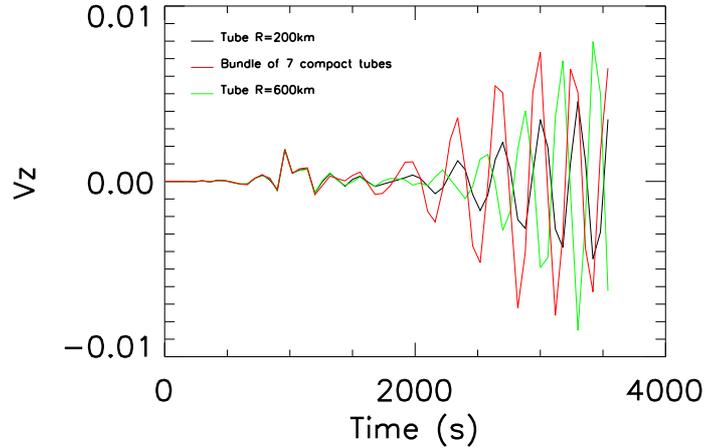} 
 \caption { Scattered vertical velocity as a function of time
   measured at point B. The black and the green curves are for monolithic flux tubes with radii of 200
   km and 600 km, respectively, situated at the same position of the central
   tube in the cluster. The red curve is the scattering from a compact cluster of seven tubes.} 
\label{seventubetime}
\end{figure}

The case of $\chi=\pi/2$ is similar to the case where the incident wave propagates
 in the $z$-direction through a pair of flux tubes. In both cases, the tubes are excited
and oscillate simultaneously.

\section{$f$-Mode Interaction with a Compact Cluster of Seven Flux Tubes} 
\label{compact7}
In this section, we study the interaction of an $f$-mode wave packet with a
cluster of seven identical magnetic flux tubes in a hexagonal
close-packed configuration. This is the simplest ``realistic'' structure that can be built
to simulate the cluster model of sunspots.

The cluster model can be a good approximation to simulate solar plage regions which are composed of an ensemble of compactly packed thin flux tubes. It is important to understand wave absorption and scattering by the plage in relation to the energy transmitted to the corona \cite{hanasoge09}.

Figure \ref{seventubetime} shows the variation of the scattered component
$V_z$ versus time at point B for a single monolithic tube of 200 km radius (black curve), a compact cluster of seven tubes (red
curve), and a single tube of 600 km radius which is the monolithic equivalent of the compact cluster (green curve).

\begin{figure} 
\centering 
\vspace{0.15\textwidth}   
\centerline{\bf      
     \hspace{0.01\textwidth}  \color{black}{(a)}
     \hfill}
\vspace{-0.15\textwidth}   
\includegraphics[width=0.65\textwidth]{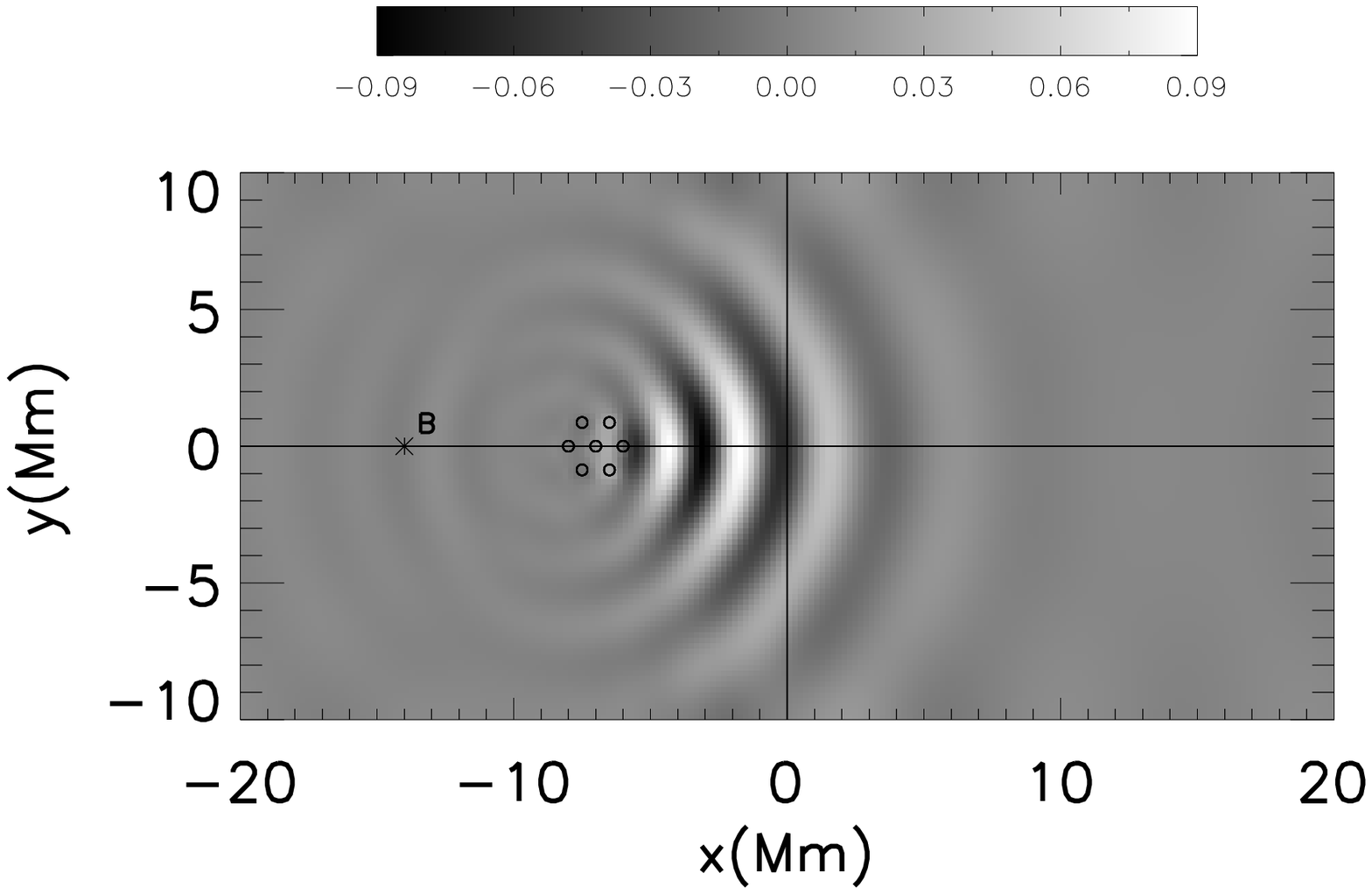}

\vspace{0.15\textwidth}  
\centerline{\bf      
     \hspace{0.01\textwidth}  \color{black}{(b)}
     \hfill}
\vspace{-0.15\textwidth} 
\includegraphics[width=0.65\textwidth]{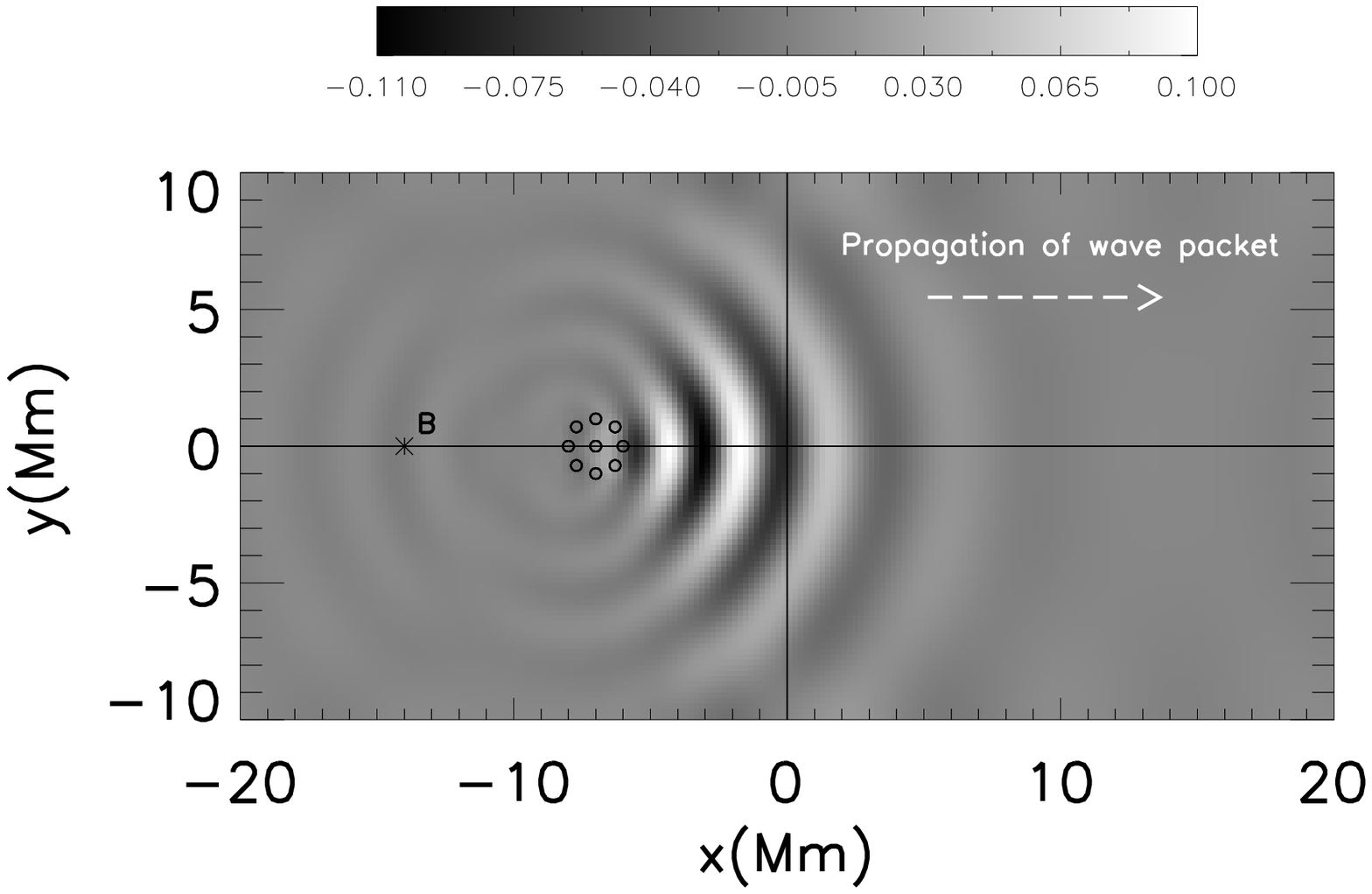}

\vspace{0.15\textwidth}  
\centerline{\bf      
     \hspace{0.01\textwidth}  \color{black}{(c)}
     \hfill}
\vspace{-0.15\textwidth} 
\includegraphics[width=0.65\textwidth]{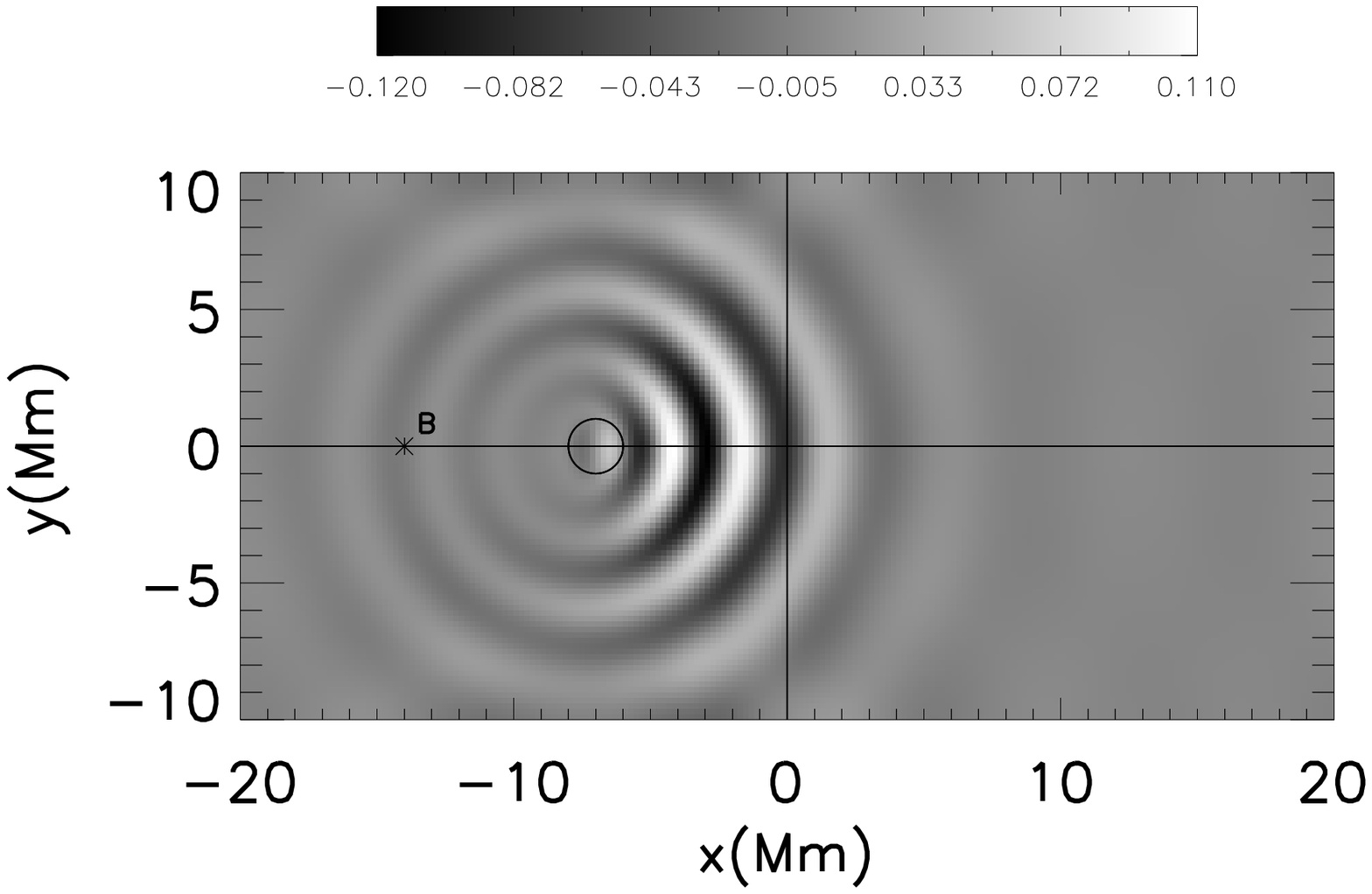} 
 \caption {(a) A snapshot at $t=3300$ s of the scattered wave field ($V_z$)
 of a loose cluster of seven identical magnetic flux tubes of 200 km
 radius. The distance from the central tube to the others is $d=0.2 \lambda$. (b) The same as (a) but for a loose  cluster of nine identical tubes of 200 km radius with separation $d=0.2 \lambda$. (c) The same as (a) but for a monolithic tube whose
radius $R=1$ Mm is the average radius of both clusters in (a) and (b).}
\label{sevenopentubetime}
\end{figure}

First, we can observe that the cluster yields a large amplitude of scattered waves 
compared to the 200 km
single tube. This amplitude is almost the same as the amplitude due to the 
 monolithic equivalent tube of 600 km radius. 

The curve of the compact cluster shows an interesting behaviour.  The cluster seems to oscillate almost like a tube of 200 km radius than like a tube of 600 km radius. 
The scattered wave fields of the compact cluster and the single tube of 200 km radius show almost an identical pattern too.
This trend may be attributed to the forcing of the collective oscillations in the compact cluster dominated by the kink mode ($m=\pm1$) in the individual flux tubes of 200 km radius, whereas the  monolithic tube of 600 km radius oscillates with a mixture of sausage ($m=0$) and kink modes \cite{bogdan96,daiffallah11}.

\section{$f$-Mode Interaction with a Loose Cluster} 

In this section, we study the scattering from a loose cluster composed of seven and nine
identical magnetic flux tubes ($R=200$ km). The distance between the central tube to the other tubes
is $d/\lambda$ = 0.2 for both clusters. 

Figure \ref{sevenopentubetime}(a)-\ref{sevenopentubetime}(c) shows the
scattered wave fields ($V_z$) of a seven-tube cluster, a nine-tube cluster, and a single tube of 1 Mm radius which is approximately the equivalent monolithic tube
to the cases of Figures \ref{sevenopentubetime}(a) and \ref{sevenopentubetime}(b).

\begin{figure}  
\centering
\includegraphics[width=0.8\textwidth]{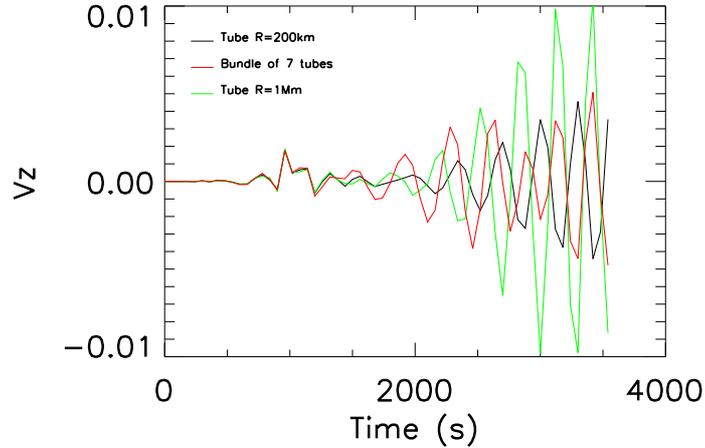} 
 \caption { Scattered vertical velocity as a function of time
   measured at point B. The black and the green curves are for monolithic flux tubes with radii of 200
   km and 1 Mm respectively, situated at the same position of the central
   tube in the cluster. The red curve is the scattering from a loose cluster of seven tubes (Figure \ref{sevenopentubetime}(a)).} 
\label{seventime}
\end{figure}

\subsection{Loose Cluster of Seven Flux Tubes}
We have done this case to compare the scattering from a compact cluster to the scattering from the same cluster when it is loosely distributed.

An important change can be seen in the near-field scattered wave to the left of the loose cluster (Figure \ref{sevenopentubetime}(a)) compared to the case of the monolithic tube. This
indicates the contribution of waves scattered by individual tubes while the compact cluster shows no multiple scattering in the near-field.

Figure \ref{seventime} shows a plot of the scattered vertical
velocity $V_z$ measured at point B as a function of time for the loose cluster of seven tubes,  and single tubes with radii of 200 km and 1 Mm. 
For separation distance $d/\lambda=0.2$, the individual tubes start to oscillate differently compared to the collective oscillations of tubes in the compact cluster; as a consequence, the oscillations of 200
km tube and the loose cluster in Figure \ref{seventime} are no longer in phase.

We found that the amplitude of scattered waves from the loose cluster is smaller compared to that from the compact cluster in Figure \ref{seventubetime}. According to the results of Section \ref{pair-simulations}, the spacing within the loose cluster ($d/\lambda$ = 0.2 from the central tube) supports only a small-amplitude oscillation of individual flux tubes which scatter waves to the the near-field (multiple scattering). Consequently, the scattering measured in the far-field will be reduced compared to that from the compact cluster since some of the incident wave energy is converted to tube oscillations. In this case, if we interpret the reduction in the far-field scattering as an enhancement of absorption of waves by the loose cluster model, then a loose cluster can be more absorbent (scatters less) than a compact cluster when $ 0.12 \lambda < d \le 0.2 \lambda$. However, when $ 0.2 \lambda < d \le 0.4 \lambda$, the loose  cluster should be more absorbent  
since oscillations of individual tubes are more efficient than in the previous case. However, coherent scattering to the far-field increases the amplitude again, in this case, measuring scattering in the far-field to evaluate absorption of waves is no longer valid.

\begin{figure}  
\centering
\includegraphics[width=0.8\textwidth]{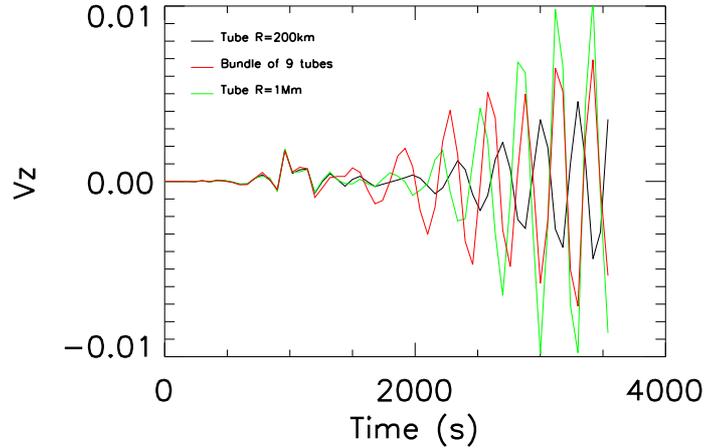} 
 \caption { A plot of the scattered vertical velocity as a function of time
   measured at point B. The black and the green curves are for monolithic flux tubes of 200
   km and 1 Mm radius, respectively, situated at the same position of the central
   tube in the cluster. The red curve is the scattering from a loose cluster of nine tubes (Figure \ref{sevenopentubetime}(b)).} 
\label{ninetubetime}
\end{figure}

\subsection{Loose Cluster of Nine Flux Tubes}

As in the previous case, multiple scattering to the near-field can be seen in the scattered wave field to the left of the loose cluster (Figure \ref{sevenopentubetime}(b)).
In Figure \ref{ninetubetime} are shown the curves of scattered vertical
velocity $V_z$ measured at point B as a function of time. From Sections \ref{pair-simulations} and \ref{compact7}, we know that the maximum scattering amplitude that can be reached by the cluster is the amplitude of the equivalent monolithic tube. From these results, it is clear that the reduction in amplitude of the loose cluster in Figure \ref{ninetubetime} is a signature of absorption of waves by individual flux tubes giving rise to the near-field scattering. We note, however, that this amplitude is larger than the amplitude of the scattered waves from the seven-tube loose  cluster. This result means that a cluster in the multiple scattering regime scatters waves more to the far-field than a less dense cluster with the same size. 

If we increase the density of the same cluster by taking $d/\lambda=0.16$, the amplitude of the scattered waves in the far-field will increase too, approaching that of the equivalent tube of 1 Mm radius.

\section{Conclusions}
While sunspots are easily visible, finding their
structure beneath the solar surface is not an easy task. Motivated by the problem of
subsurface magnetic structure of sunspots, we investigate numerically the wave
propagation through a cluster model of magnetic flux tubes. The goal is to
distinguish helioseismically between this model and the monolithic model of
sunspots by observing the scattered wave field.  

The multiple flux tube model is a good model for plage as well. The attenuation
of the waves in the plage regions is enhanced and this type of simulation
provides a good way of understanding that.

In the first part of this study, we have simulated the
interaction between an $f$-mode wave packet and a pair of small identical magnetic
flux tubes of radius $R$ = 200 km positioned along ($x$-direction) or perpendicular ($y$-direction) to the direction of propagation of the incoming waves.

For the pair aligned in the $x$-direction, when the distance between the tubes $d$ is less than $\lambda/2\pi$
 ($\lambda$ is the wavelength of the incident wave packet), the pair of tubes oscillate as a single tube. The result
is a maximum amplitude of scattering measured at the far-field.
The amplitude of the scattered wave from the compact pair is about the same as that from a single tube of $2R$ radius. 

When separation $d$ is about $\lambda/2\pi$, the individual tubes start to oscillate and scatter waves to the
near-field (multiple scattering regime) taking a part of
the energy of the scattered wave. Then, the amplitude of the scattered wave decreases. 
When $d$ is about twice of $\lambda/2\pi$, oscillations from
individual tubes increase and reach the level of the far-field scattering, giving rise to a coherent
scattering. The amplitude increases again to about that from  the compact
pair of tubes (coherent scattering regime). 

For the pair of flux tubes aligned in the $y$-direction, the tubes oscillate
simultaneously with the incoming waves.

In order to obtain more realistic models for sunspots and plage, we have invesigated the
propagation of waves through a cluster of small identical magnetic flux
tubes of 200 km radius. We have studied
two cases, one is a compact cluster and the other is a loose cluster. The compact
cluster seems to oscillate more like a single tube of 200 km radius than like the monolithic equivalent tube, while the scattering behavior
of the loose cluster shows multiple-scattering from the individual tubes in the
near-field. However, the scattered amplitude of the compact cluster measured in the far-field is almost the same as the
scattered amplitude of the monolithic equivalent tube.

We have demonstrated that a loose cluster in the multiple scattering regime is more efficient in absorption of waves than a compact cluster or the equivalent monolithic tube of both kinds of clusters.  It is reasonable to infer that less is the density of tubes within a cluster, more is the absorption of waves by individual flux tubes and less is the scattering to the far-field. However, it is unclear how to evaluate the absorption of waves by a cluster in the coherent scattering regime since scattering is enhanced in the far-field.

Notice that we are discussing interaction of surface gravity waves ($f$-mode) with small radius flux tubes. In this context, the scattering $f$-$f$ modes will be the predominant process compared to the absorption and propagation of waves in the $z$-direction. Therefore, it will be no surprising to find more absorption of waves for $p$-mode interaction with magnetic flux tubes and clusters.

Future investigations require more simulations using different sizes of
magnetic flux tubes, various geometrical configurations, $p$-modes as an incident wave packet.
More spatial and temporal resolutions are required to study the near-field phenomena and the interaction among tubes within the cluster.


\bibliographystyle{spr-mp-sola}

\bibliography{sola_bibliography_example}  

\IfFileExists{\jobname.bbl}{} {\typeout{}
\typeout{****************************************************}
\typeout{****************************************************}
\typeout{** Please run "bibtex \jobname" to obtain} \typeout{**
the bibliography and then re-run LaTeX} \typeout{** twice to fix
the references !}
\typeout{****************************************************}
\typeout{****************************************************}
\typeout{}}

\end{article} 

\end{document}